\documentclass[twocolumn, tighten, dvipsnames, floatfix]{aastex63}

\usepackage[]{xcolor}
\usepackage{booktabs}
\usepackage{physics}
\usepackage{longtable}
\usepackage{array}
\usepackage{changepage}
\usepackage{lineno}
\usepackage[flushmargin]{footmisc}

\accepted{\normalsize for ApJ}

\usepackage{lipsum}
\usepackage{float}

\shorttitle{Dynamically Tagged Groups of Very Metal-poor HK/HES Halo Stars}
\shortauthors{Limberg et al.}

\begin{document}

\title{Dynamically Tagged Groups of Very Metal-poor Halo Stars from the HK and Hamburg/ESO Surveys} 

\correspondingauthor{Guilherme Limberg}
\email{guilherme.limberg@usp.br}

\author[0000-0002-9269-8287]{Guilherme Limberg}
\affil{Universidade de S\~ao Paulo, Instituto de Astronomia, Geof\'isica e Ci\^encias Atmosf\'ericas, Departamento de Astronomia, SP 05508-090, S\~ao Paulo, Brazil}

\author[0000-0001-7479-5756]{Silvia Rossi}
\affiliation{Universidade de S\~ao Paulo, Instituto de Astronomia, Geof\'isica e Ci\^encias Atmosf\'ericas, Departamento de Astronomia, SP 05508-090, S\~ao Paulo, Brazil}

\author[0000-0003-4573-6233]{Timothy C. Beers}
\affiliation{Department of Physics and JINA Center for the Evolution of the Elements, University of Notre Dame, Notre Dame, IN 46556, USA}

\author[0000-0002-0537-4146]{H\'elio D. Perottoni}
\affiliation{Universidade de S\~ao Paulo, Instituto de Astronomia, Geof\'isica e Ci\^encias Atmosf\'ericas, Departamento de Astronomia, SP 05508-090, S\~ao Paulo, Brazil}

\author[0000-0002-5974-3998]{Angeles P\'erez-Villegas}
\affiliation{Universidade de S\~ao Paulo, Instituto de Astronomia, Geof\'isica e Ci\^encias Atmosf\'ericas, Departamento de Astronomia, SP 05508-090, S\~ao Paulo, Brazil}
\affiliation{Instituto de Astronom\'ia, Universidad Nacional Aut\'onoma de M\'exico, Apartado Postal 106, C. P. 22800, Ensenada, B. C., M\'exico}

\author[0000-0002-7529-1442]{Rafael M. Santucci}
\affiliation{Universidade Federal de Goi\'as, Instituto de Estudos S\'ocio-Ambientais, Planet\'ario, Goi\^ania, GO 74055-140, Brazil}
\affiliation{Universidade Federal de Goi\'as, Instituto de F\'isica, Goi\^ania, GO 74001-970, Brazil}

\author[0000-0002-6838-2178]{Yuri Abuchaim}
\affiliation{Universidade de S\~ao Paulo, Instituto de Astronomia, Geof\'isica e Ci\^encias Atmosf\'ericas, Departamento de Astronomia, SP 05508-090, S\~ao Paulo, Brazil}

\author[0000-0003-4479-1265]{Vinicius M. Placco}
\affiliation{NSF's Optical-Infrared Astronomy Research Laboratory, Tucson, AZ 85719, USA}

\author[0000-0003-0852-9606]{Young Sun Lee}
\affiliation{Department of Astronomy and Space Science, Chungnam National University, Daejeon 34134, Korea}

\author[0000-0002-4043-2727]{Norbert Christlieb}
\affiliation{Zentrum f$\ddot{u}$r Astronomie der Universit$\ddot{a}$t Heidelberg, Landessternwarte, K$\ddot{o}$nigstuhl, 69117 Heidelberg, Germany}

\author[0000-0002-7900-5554]{John E. Norris}
\affiliation{Research School of Astronomy and Astrophysics, Australian National University, Canberra, ACT 2611, Australia}

\author[0000-0001-7801-1410]{Michael S. Bessell}
\affiliation{Research School of Astronomy and Astrophysics, Australian National University, Canberra, ACT 2611, Australia}

\author[0000-0001-9069-5122]{Sean G. Ryan}
\affiliation{School of Physics, Astronomy and Mathematics, University of Hertfordshire, College Lane, Hatfield AL10 9AB, UK}

\author[0000-0002-4792-7722]{Ronald Wilhelm}
\affiliation{Department of Physics and Astronomy, University of Kentucky, Lexington, KY 40506, USA}

\author[0000-0001-9214-7437]{Jaehyon Rhee}
\affiliation{Center for Astrophysics $\vert$ Harvard~\&~Smithsonian, 60 Garden Street, MS-09, Cambridge, MA 02138, USA}

\author[0000-0002-2139-7145]{Anna Frebel}
\affiliation{Department of Physics and Kavli Institute for Astrophysics and Space Research, Massachusetts Institute of Technology, Cambridge, MA 02139, USA}

\bigskip\bigskip\smallskip\smallskip\smallskip\smallskip\smallskip\smallskip\smallskip

\begin{abstract}

We analyze the dynamical properties of $\sim$1500 very metal-poor (VMP; [Fe/H] $\lesssim -2.0$) halo stars, based primarily on medium-resolution spectroscopic data from the HK and Hamburg/ESO surveys. These data, collected over the past thirty years, are supplemented by a number of calibration stars and other small samples, along with astrometric information from $Gaia$ DR2. We apply a clustering algorithm to the 4-D energy-action space of the sample, and identify a set of 38 Dynamically Tagged Groups (DTGs), containing between 5 and 30 member stars.  Many of these DTGs can be associated with previously known prominent substructures such as \textit{Gaia}-Sausage/Enceladus (GSE), Sequoia, the Helmi Stream (HStr), and Thamnos. Others are associated with previously identified smaller dynamical groups of stars and streams. We identify 10 new DTGs as well, many of which have strongly retrograde orbits. We also investigate possible connections between our DTGs and $\sim$300 individual $r$-process-enhanced (RPE) stars from a recent literature compilation. We find that several of these objects have similar dynamical properties to GSE (5), the HStr (4), Sequoia (1), and Rg5 (1), indicating that their progenitors might have been important sources of RPE stars in the Galaxy. Additionally, a number of our newly identified DTGs   are shown to be associated with at least two RPE stars each (DTG-2: 3, DTG-7: 2; DTG-27: 2). Taken as a whole, these results are consistent with ultra-faint and/or dwarf spheroidal galaxies as birth environments in which $r$-process nucleosynthesis took place, and then were disrupted by the Milky Way.

\end{abstract}


\section{Introduction} 
\label{sec:intro}

The currently accepted model for the formation of the Galactic stellar halo (hereafter ``halo") consists of frequent mergers between the nascent Milky Way (MW) and dwarf satellite galaxies of various masses, based on early suggestions from \citet{sz1978}, and numerous efforts since. This bottom-up scenario is supported by theoretical predictions from the $\Lambda$ Cold Dark Matter cosmological paradigm \citep{spergel2007}, and numerical simulations of increasing sophistication based on it (see, e.g., \citealt{somerville2015}).   

\setcounter{footnote}{12}

The discovery of the Sagittarius dwarf spheroidal, a galaxy in the process of tidal disruption \citep{ibata1994}, and smaller stellar streams in the halo (e.g., \citealt{helmi1999, chiba2000}) provided strong evidence for this assembly mechanism. Metallicities ([Fe/H]\footnote{Definition of abundance of a star ($\star$) relative to the Sun ($\odot$): [A/B] $= \log \rm (N_A/N_B)_\star - \log (N_A/N_B)_\odot$, where $\rm N_A$ ($\rm N_B$) are the number densities of atoms for elements A (B).}), radial velocities, and proper motions made available by large surveys such as the Sloan Digital Sky Survey \citep[SDSS;][]{sdssYork}, in particular its stellar-specific sub-survey Sloan Extension for Galactic Exploration and Understanding (SEGUE: \citealt{yanny2009}), led to the proposition (supported by many studies since) that ``the halo" comprises at least two overlapping stellar populations, the inner-halo population (IHP) and the outer-halo population (OHP), with differences in their spatial density distributions, stellar kinematics, and chemical abundances \citep{carollo2007, carollo2010, dejong2010, beers2012,an2013,an2015, lee2017, lee2019,kim2019, an2020}.

Even limited kinematic information has been used to infer the presence of substructure throughout the halo of the MW.  For example, \citet{schlaufman2009,schlaufman2011,schlaufman2012} used radial velocities alone to identify a plethora of ECHOS (Elements of Cold Halo Substructures) within the inner-halo region from the SDSS/SEGUE surveys. \citet{an2013,an2015} combined proper motions with photometric metallicities to estimate the fractions of IHP and OHP stars in the local neighborhood. \citet{an2020} used the powerful combination of photometric metallicities and precision astrometric surveys to produce a ``blueprint" of the known stellar populations in the disk and halo systems of the Galaxy. In the process, these authors confirmed the presence of a dynamically heated disk population, named the ``Splashed Disk" (SD; \citealt{bonaca2017, DiMatteo2019, belokurov2020, amarante2020a, Amarante2020b}), and demonstrated that the metal-weak thick disk (MWTD) is an independent structure from the canonical thick disk (see also \citealt{carollo2019}).

Many recent works have used full space motions, based on radial velocities and astrometry (parallaxes and proper motions) provided by \textit{Gaia}'s Data Releases (DRs; \citealt{GaiaMission,gaiadr1, gaiadr2}), combined with previously available spectroscopic and photometric data, to analyze the kinematics and abundances for very large samples of halo and disk stars (see \citealt{helmi2020} for a review). It has been proposed that the IHP is dominated by the remnant of a single, relatively massive (stellar mass $M_{\star} \sim 6 \times 10^8 M_{\odot}$) merger event, some 10 Gyr ago, named the \textit{Gaia}-Sausage \citep{belokurov2018,Haywood2018, MyeongSausageClusters} or \textit{Gaia}-Enceladus \citep{helmi2018}. Orbital modeling indicates that stars in this system present highly radial orbits (a signature noted very early on by, e.g., \citealt{norris1986, sommer-larsen1997,chiba2000, ryan2003}). In an independent effort, \citet{helmi2018} investigated the properties of this substructure, and were able to associate it with one of the two distinct main-sequence turnoffs identified in the local halo \citep{gaiaHR}.

In a series of papers, \citet{myeongHalo, myeongShards, myeongSequoia} argued that a population of high orbital energy (hereafter ``energy"; $E$), retrograde halo stars partially overlaps with \textit{Gaia}-Enceladus, differentiating this proposed event from the one described by \citet{belokurov2018}. Indeed, \citet{myeongSequoia} suggested that a different substantial progenitor could be attributed to this substructure, the Sequoia galaxy ($M_\star \sim 5 \times 10^7 M_{\odot}$), providing the bulk of the high-energy, retrograde outer-halo stars. On the other hand, \citet{koppelman2019} attributed the low-energy counterpart of Sequoia to a different merging event of smaller scale, named Thamnos ($M_\star \sim 5 \times 10^6 M_{\odot}$), previously reported in part by \citet{Helmi2017} and \citet{koppelman2018}. Stars linked to this system also have  higher values of both [Mg/Fe] and [Al/Fe], thus suggesting a chemical separation from Sequoia as well. In contrast to \textit{Gaia}-Sausage/Enceladus (GSE), Thamnos exhibits lower values of [Fe/H], consistent with the progenitor being a smaller galaxy.

All of the above efforts have contributed greatly to our current understanding of the complex formation history of the Galactic halo, focusing on identifying its most prominent substructures. However, low-mass dwarf galaxies accreted by the MW and disrupted into the halo would not be expected to present strong spatial over-densities (streams), in particular in the low stellar-density outer-halo region. For the purpose of finding the remnants of such systems, one would consequently want to construct samples of stars with similar characteristics to their parent mini-halos' stellar populations, increasing the fraction of objects that originated in these environments (see \citealt{simon2019} for a review).

Since low-mass dwarf satellite galaxies primarily host very metal-poor (VMP; [Fe/H] $<-2.0$) stars, \citet{yuan2020a} proposed the examination of stellar samples of VMP stars in order to identify their debris in the local halo. These authors applied a neural-network based technique to dynamically cluster their sample of $\sim$3000 stars (from LAMOST DR3; \citealt{lamostVMP}, with re-determined stellar parameters by Beers) in energy and angular-momentum space. They found 57 Dynamically Tagged Groups (DTGs), many of them related to larger substructures, such as GSE, and some groups from \citet{myeongShards}. They were also able  to associate previously known chemically peculiar ($r$-process-enhanced or carbon-enhanced) stars with them. The pioneering effort of \citet{roederer2018} identified eight dynamical groups of $r$-process-enhanced (RPE) stars (with 3 or 4 stars each) from a relatively small sample of 35 such objects. Chemo-Dynamically Tagged Groups have been identified from much larger samples of RPE and carbon-enhanced metal-poor (CEMP; [C/Fe] $> +0.7$ and [Fe/H] $< -1.0$) stars by \citet{gudin2020a} and Gudin et al. (2020, in preparation), respectively. Furthermore, \citet{yuan2020b} used metal-poor blue horizontal-branch and RR Lyrae stars to discover a low-mass stellar-debris stream apparently associated with a pair of globular clusters in the outer-halo region, which they named LMS-1. This substructure was independently confirmed by \citet{naidu2020}, which they called Wukong. There are surely many more such small groups/streams that remain to be identified. 

VMP substructures recognized from these efforts are possibly remnants of ultra-faint dwarf (UFD) and low- to intermediate-mass dwarf spheroidal (dSph) galaxies that were accreted and shredded by the MW. Since there now exists a plentiful supply of VMP stars that are much closer (and hence brighter) than individual stars in any surviving dwarf satellite, these objects provide an opportunity to study the chemical-evolution histories of their now-disrupted parent systems in much greater detail. This possibility is particularly appealing in the context of the recent discoveries of RPE stars in the UFD galaxies Reticulum II \citep{ji2016,roed2016} and Tucana III \citep{hansen2017, marshall2019}. Classical dSph galaxies might also present moderately enhanced ($r$-I; $+0.3 < $ [Eu/Fe] $ \leq +0.7$) and highly enhanced ($r$-II; [Eu/Fe] $ > +0.7$)\footnote{We adopt the definitions for $r$-process enrichment from \citet{holmbeck2020}.} RPE stars. 

\begin{figure*}[t!]
\centering
\includegraphics[scale=0.4]{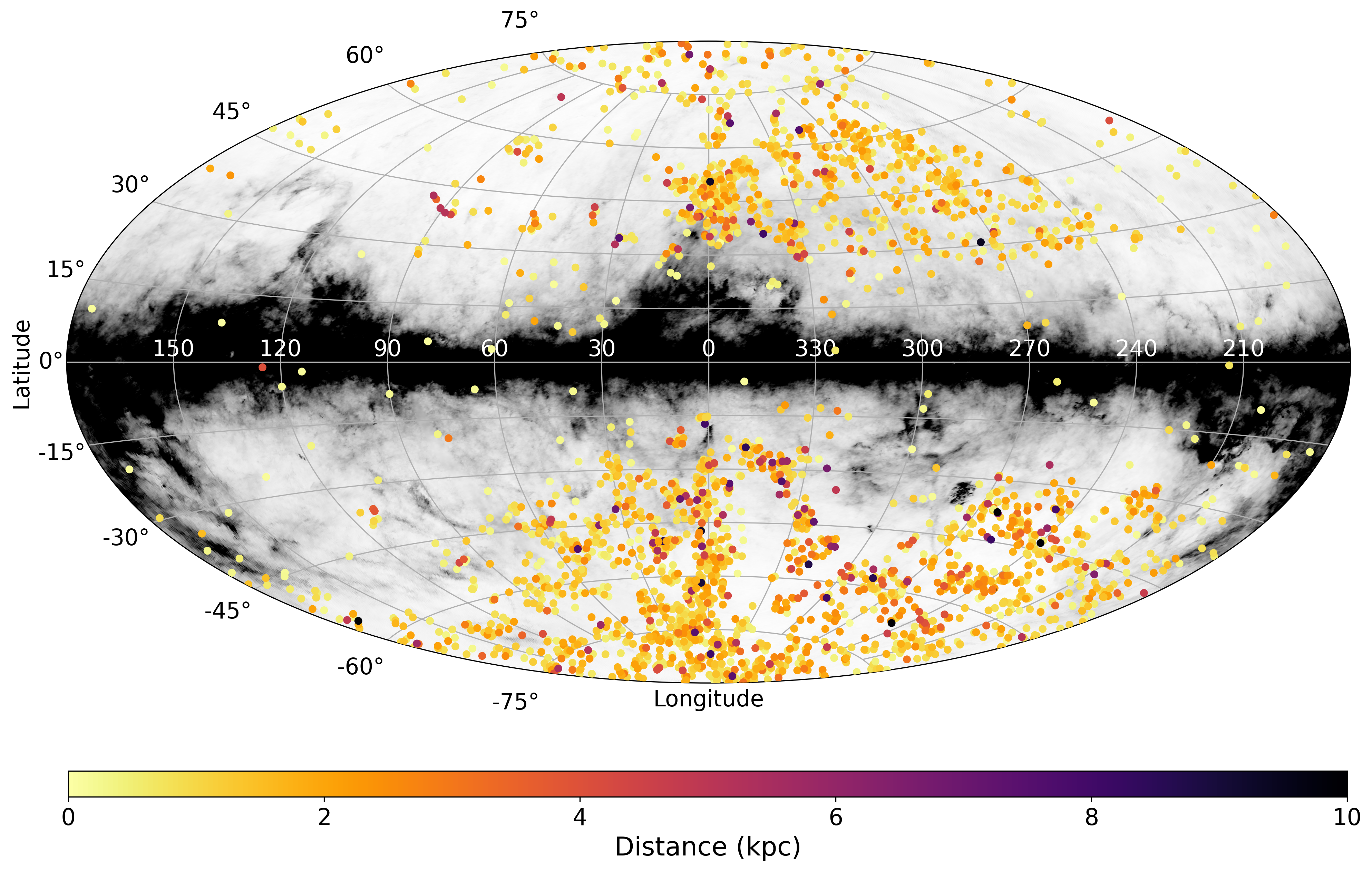}
\caption{Distribution of the VMP HK/HES sample in the Galactic coordinate system, color-coded by heliocentric distances (see text) with relative errors smaller than 20\% of the nominal values (Section \ref{sec:data}). The background all-sky distribution of the Galactic reddening comes from the \citet{ExtictionMap1} map, as re-calibrated by \citet{ExtictionMap2}. The different gray scales represent E$(B-V)$ values from 0.0 (white) to 0.5 (black).
\label{fig:positions}} 
\end{figure*}

The fundamental goal of this paper is to identify additional  fragments of dwarf satellites that have been merged with the Galactic halo. We also investigate possible associations of such systems with the large sample of RPE stars compiled by \citet{gudin2020a}. We establish a straightforward, easily reproducible framework to identify DTGs, and apply it to the combined sample of some 1500 VMP stars (with available estimates of metallicity, radial velocities, and astrometric data, after removal of possible non-halo stars and stars with uncertain distance estimates) originally identified in the HK survey of Beers and colleagues and the Hamburg/ESO (HES) survey of Christlieb and collaborators, along with a number of calibration stars and halo star candidates with spectroscopic data taken during the HK/HES surveys (see Figure~\ref{fig:positions}).  The most interesting of these DTGs will provide the opportunity to explore the nucleosynthetic processes operating in the environments of UFD and/or dSph galaxies in the past. 

This paper is organized as follows. Section~\ref{sec:data} describes the assembly of our VMP sample, including estimates of their stellar atmospheric parameters, as well as their kinematic and astrometric data. Section~\ref{subsec:dyn} reports our calculations of the dynamical properties for the stars in this sample. Section~\ref{sec:subs} describes our substructure search methodology, and the assignment of individual stars into 38 DTGs. Our analysis of these DTGs is presented in Sections \ref{sec:large} and \ref{sec:new}, including their possible associations with previously recognized substructures and dynamical groups. In Section \ref{subsec:$r$-process}, we map previously known RPE stars onto our DTGs. Finally, Section \ref{sec:conclusion} provides concluding remarks and a brief discussion.

\section{Data} \label{sec:data}

\subsection{The HK/HES Spectra}

Beginning almost fifty years ago with the pioneering work of \citet{bond1970, bond1980} and \citet{bidelman1973}, photographic objective-prism techniques have proven to be efficient sieves for identifying large numbers of stars that are metal deficient (and/or chemically peculiar). These efforts were expanded by the HK survey\footnote{The first observations on this program were made in the late 1970's by George Preston and Stephen Shectman, hence in the early literature this survey is sometimes referred to as the ``Preston-Shectman" survey.} \citep{beers1985,beers1992}, and later by the HES survey \citep{chris2008}, which included fainter stars, making it possible to explore deeper into the Galactic halo, where more metal-poor stars have been found. Both surveys sought to identify such stars via visual (HK) or automated (HES) inspection of the prism plates, searching for stars with weak, or absent, Ca~II K lines in their very low-resolution ($R = \lambda/\Delta\lambda \sim 300$) spectra.

Over the past three decades, the metal-poor candidates identified in these surveys have been followed-up with medium-resolution ($1200 \lesssim R \lesssim 2000$) spectroscopy with a wide variety of telescopes and instruments.  The typical spectral coverage of these spectra are 3500-5000\AA{}, although there was wide variation. The decisions as to which targets to observe are difficult to quantify, since many observers contributed to these efforts.  In addition, in some cases, photometric information from independent efforts (HK), or taken directly from an approximate calibration of the prism spectra (HES), were obtained in advance of the spectroscopic follow-up. A complete description of the HK and  HES candidate selection can be found in \citet{beers1985,beers1992} and \citet{chris2008}, respectively. 

Metallicity ([Fe/H]) estimates were originally obtained by application of a number of techniques, initially based solely on the indices tracking the equivalent width of the Ca~II K line, as a function of measured or estimated color (often $B-V$; see \citealt{beers1990b}). Later,  techniques designed to obtain estimates of the carbonicity ([C/Fe]), based on the strength of the CH $G$-band feature at $\sim$4330\AA{}, were developed (see, e.g., \citealt{rossi2005,placco2010,placco2011}, and references therein). Further refinements, including approaches to  avoid difficulties involving the saturation of Ca~II K indexes at higher metallicities and/or lower temperatures were incorporated (see, e.g., \citealt{beers1999, rossi2005}). Many of these same techniques were used as the starting point for more modern pipelines used for similar medium-resolution spectra from SDSS \citep{lee2008a, lee2008b} and LAMOST \citep{xiang2015}. 

\begin{figure}[htp]
\centering
\includegraphics[width=\columnwidth]{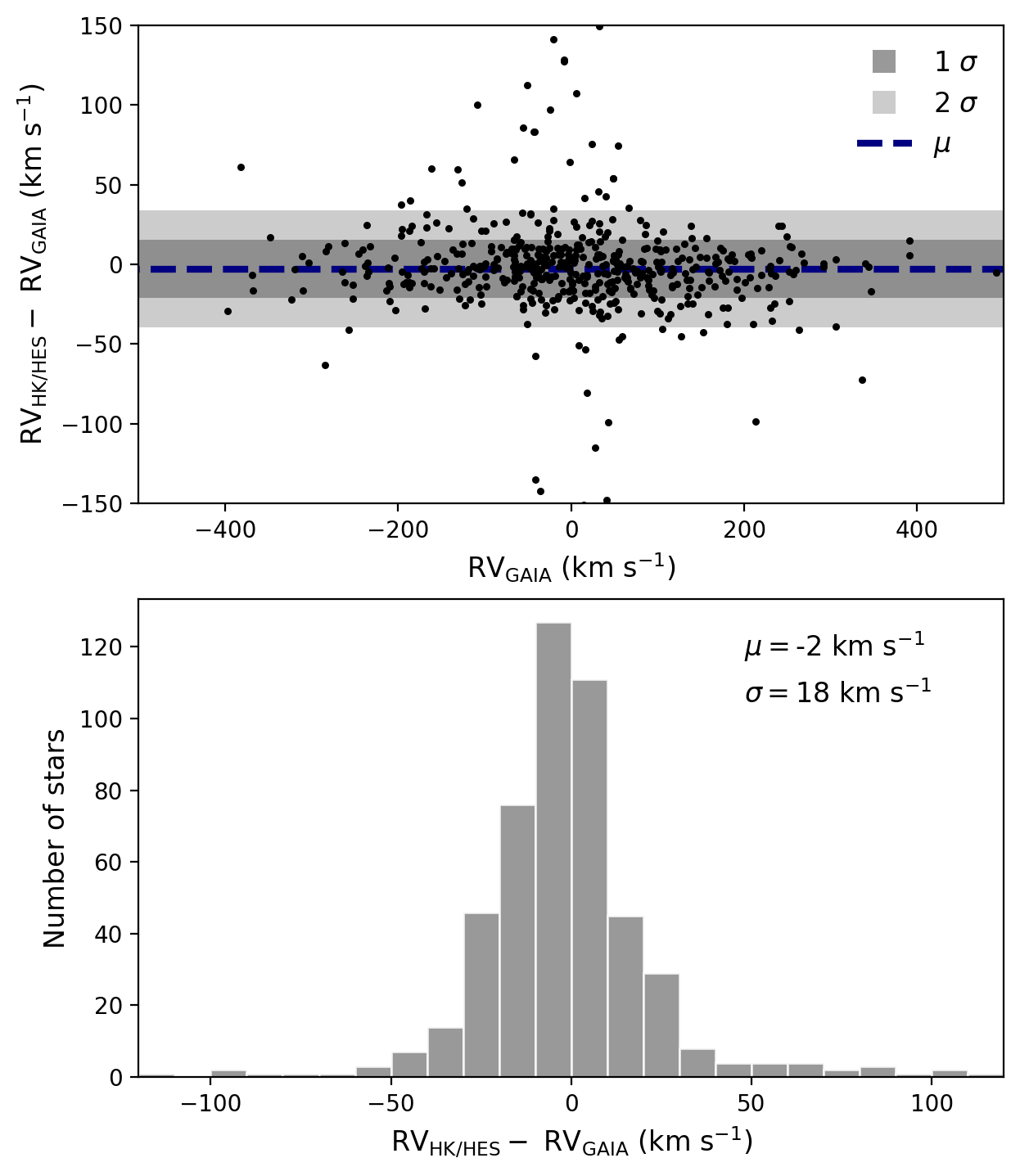}
\caption{Upper panel: Residuals between measured (line-by-line or cross-correlation) RVs of the VMP HK/HES sample and those from \textit{Gaia} DR2. The shaded areas represent the $1\sigma$ and $2\sigma$ ranges, where $\sigma$ is the biweight scale. The dashed blue line represents the biweight central location ($\mu$). Lower panel: Histogram of the residuals.}
\label{fig:RV_comp}
\end{figure}

One of the primary drivers for these endeavors was to develop lists of vetted VMP stars that served as input targets for large-scale high-resolution spectroscopic follow-up campaigns involving astronomers worldwide (e.g., the ``Extremely Metal-poor Stars'' program of Norris et al., e.g., \citealt{norris1996}; the ``First Stars'' program of Cayrel et al., e.g., \citealt{hill2002,cayrel2004}; the CEMP stars follow-up program by Aoki et al., e.g., \citealt{aoki2007}; the ``Zero-Z'' program of Cohen et al., e.g., \citealt{cohen2008}).  For almost two decades, the HK/HES surveys were responsible for the discovery of the majority of stars known with [Fe/H] $< -3.0$ (see \citealt{frebel2006, placco2011, roederer2014}), including the first hyper metal-poor ([Fe/H] $< -5.0$) stars found in the Galactic halo \citep{chris2002,frebel2005}.

\subsection{Re-Analysis of the HK/HES Spectra}

As the determinations of [Fe/H] (and [C/Fe]) for the HK/HES medium-resolution spectra were acquired over many years (some including photometric information, some not), with a variety of instruments and calibrations, for our present purpose it is necessary to perform a homogeneous re-analysis. Of course, we now have available multiple new techniques, photometry, and high-resolution calibrations that can be brought to bear on this effort.

All estimates of stellar atmospheric parameters for our stars have been obtained by application of the n-SSPP pipeline \citep{beers2014,beers2017}, a modified version of the SEGUE Stellar Parameter Pipeline (SSPP; \citealt{lee2008a,lee2008b,lee2011,lee2013}). The n-SSPP is a compilation of routines that utilizes spectroscopic and photometric inputs to perform various estimates of the stellar parameters. It employs $\chi^2$ minimization between the analysed spectra and a dense grid of synthetic ones, as well as other techniques, where suitable, depending on the wavelength coverage of the input. Then, the best set of values is adopted. See also \citet{placco2018, placco2019} for recent applications of these methods to low-metallicity stars observed with a variety of instruments. The errors for effective temperatures ($T_{\rm eff}$) and surface gravity ($\log g$) values are $\pm$150 K and $\pm$0.35 dex, respectively. The adopted solar abundances are from \citet{asplund2009}. The typical uncertainty for estimates of [Fe/H] and [C/Fe] is $\pm$0.15-0.25 dex, depending on $T_{\rm eff}$ and signal-to-noise ratio. Comparisons between the metallicity and carbonicity values from the n-SSPP and those from high-resolution spectroscopy are published in \citet{placco2014}; they are quite compatible (at the $1\sigma$ level).

The radial velocities (RVs) have been measured with the line-by-line and cross-correlation techniques (\citealt{beers1999}; see also \citealt{beers2014, beers2017}) and are precise to 10-15\,km\,s$^{-1}$. Figure \ref{fig:RV_comp} shows the comparison between our measured RVs and those from \textit{Gaia} DR2 (when available). The shaded areas represent the $1\sigma$ and $2\sigma$ ranges, where $\sigma$ is the biweight scale \citep{beers1990a}. This metric is more suitable, since the distribution of residuals is affected by occasional outliers. Many of the larger differences are likely due to the presence of binary systems or problems with the wavelength calibration of the medium-resolution spectra for individual stars. We remove any stars residing outside this $2\sigma$ region from our analysis, whenever \textit{Gaia} RVs are available for comparison. Even though other sources of RVs are usable for part ($\sim$20\%) of our sample, we have decided to retain these directly measured values in order to preserve the homogeneity of our uncertainties, and avoid introducing unnecessary biases.

\subsection{The VMP HK/HES Sample}
\label{subsec:hkhesVMPsample}

To construct our Initial Sample, we first remove apparent white dwarfs or other warm objects, such as subdwarf B stars, spectra with clear Ca~II K-line core emission features, or spectral defects in the wavelength region of the Ca~II K/H lines (3900\,{\AA}-4000\,{\AA}). For objects with more than a single observation, we adopt the median of the estimated metallicities. We then exclude anything bluer than the main-sequence turnoff by limiting our sample to $T_{\rm eff} < 7000$\,K. Note that this temperature cutoff still includes a number of halo blue stragglers and horizontal-branch stars. Finally, we have selected objects with [Fe/H] $\leq-1.8$, as these will include VMP stars within the expected error bars on the metallicity estimates. We refer to all of these stars below as VMP stars. This results in a total Initial Sample of 4,443 VMP stars.

\begin{figure*}[ht!] 
\centering
\includegraphics[width=18.0cm]{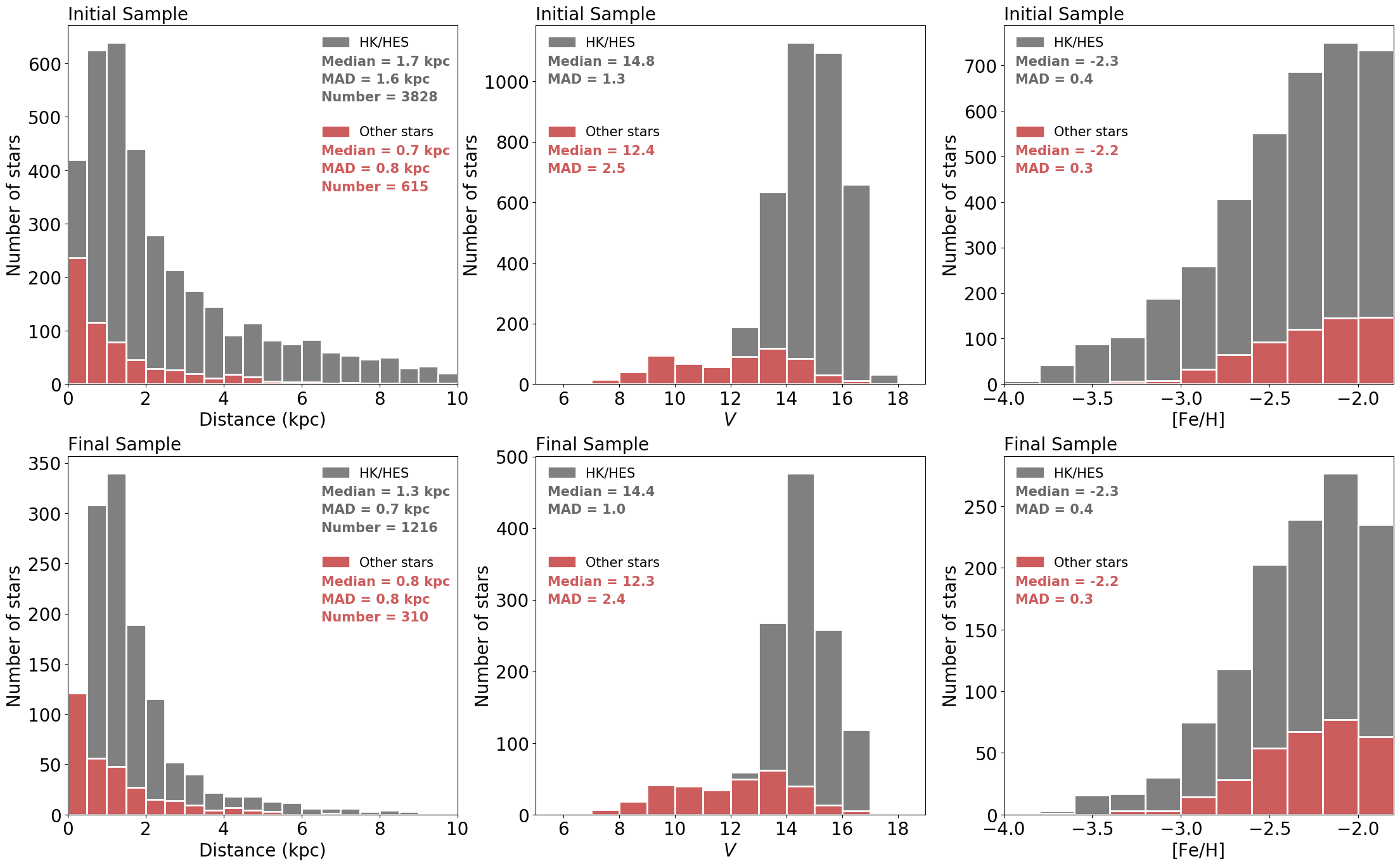}
\caption{Histograms of the distributions of heliocentric distances, $V$-band magnitudes, and [Fe/H] for both the Initial Sample (upper row) and the Final Sample (lower row), as described in Section \ref{sec:data}. Gray bars represent the VMP HK/HES stars. Red bars represent stars from other sources, including many calibration stars, as discussed in Section \ref{subsec:hkhesVMPsample}. Medians and median absolute deviation (MAD) values have been listed following the color scheme of the histograms. The total number of stars of each sample has been added in the left panels.}
\label{fig:hists}
\end{figure*}

Note that the Initial Sample includes a small fraction ($\sim$15\%) of stars -- primarily used for parameter calibration and/or short lists of candidate halo stars from other samples -- that were observed with the same resolving power and wavelength coverage as the HK/HES stars during the course of the main follow-up effort (see Figure \ref{fig:hists}). These other small samples include data from: ($i$) \citet{beers2001}, comprising cooler stars from the Edinburgh-Cape Blue Object Survey (indicated by the prefix `EC' in our sample); ($ii$) \citet{beers2002}, comprising candidate metal-poor stars identified close to the Galactic plane in the Luminous Stars Extension survey (identified with the prefix `LSE' in our sample); ($iii$) \citet{rhee2001}, comprising metal-poor candidates (primarily cooler giants) identified from digital scans of the HK survey plates with the Automatic Plate Machine facility in Cambridge, UK \citep{rhee1999}, known as the HK II survey (identified with the prefix `II' in our sample); ($iv$) \citet{frebel2006} (see also \citealt{beers2017}), comprising bright metal-poor candidates from the HES plates; and ($v$) \citet{beers2014}, comprising metal-poor candidates selected by \citet{bidelman1973} (see also \citealt{norris1985}). For simplicity of notation, we refer to the full sample as the VMP HK/HES sample (listed in the Appendix \ref{sec:append_Init}; Table \ref{tab:initial_sample}).

We cross-match the VMP HK/HES sample with \textit{Gaia} DR2 to obtain accurate proper motions, where available. We combine these data with distances from \citet{anders2019}, estimated with the \texttt{StarHorse} code \citep{queiroz2018, queiroz2020} in a Bayesian framework. The 50th percentile of these distance distributions are compatible with the inverse of the \textit{Gaia} parallaxes within 2-3 kpc from the Sun, but can be used when the reported $Gaia$ parallax is either negative or missing. Finally, we have restricted our sample to stars with relative distance errors smaller than 20\% of their nominal values, assuming Gaussian distributions according to its 16th and 84th percentiles. This cut yields heliocentric distances $\lesssim$ 5 kpc (Figure \ref{fig:hists}), with few exceptions. Out of these stars, the vast majority ($\sim$97\%) have re-normalized unit weight errors within the recommended interval (\texttt{RUWE} $< 1.4$; \citealt{lindegren2018}), which can be used to obtain reliable dynamical-parameter estimates.

We have adopted a velocity of the Local Standard of Rest (LSR) of $V_{\rm{LSR}}=232.8$\,km\,s$^{-1}$ \citep{mcmillan2017}, and a peculiar motion of the Sun with respect to the LSR of ($U$,$V$,$W$)$_{\odot}~=~(11.1,12.24,7.25)$\,km\,s$^{-1}$ \citep{schon2010}. Then, the position on the sky, distance, proper motions, and RVs of the stars are converted to the Cartesian Galactic phase-space positions and velocities using {\tt Astropy} Python tools \citep{astropy, astropy2018}. Finally, we have made a cut in velocity, $|V-V_{\rm{LSR}}|>210$\,km\,s$^{-1}$, to primarily retain stars from the halo. Applying this criterion leaves a Final Sample of 1,540 likely VMP HK/HES halo stars with data suitable for dynamical analysis (these are listed, along with the adopted dynamical parameters, in the Appendix \ref{sec:append_Final}; Table \ref{tab:final_sample}).

\subsection{The RPE Stars Sample}
\label{subsec:rpeSample}

For the mapping of RPE stars onto our DTGs, carried out in Section \ref{subsec:$r$-process}, we have adopted the recent compilation of $r$-I and $r$-II stars from \citet{gudin2020a}. The majority of these objects come from the $R$-Process Alliance data releases \citep{hansen2018, sakari2018, ezzedine2020, holmbeck2020}, complemented with additional data from JINAbase \citep{jina}. These stars are all metal-poor ([Fe/H] $<-1.0$) and at least moderately enriched in $r$-process elements ([Eu/Fe] $> +0.3$; [Ba/Eu] $<0.0$). We do not apply the [Fe/H] $\leq -1.8$ cut since these stars are already chemically peculiar, independently of their metallicities. For consistency, we have used the same kinematic criteria described in Section \ref{subsec:hkhesVMPsample}, but keeping stars with relative distance errors up to 30\%, which avoids constraining our sample too much. These larger errors in distances lead to increasing uncertainties in the dynamical parameters. However, since we employ a statistical method for the cluster assignment, such errors will propagate into smaller membership probabilities, which we independently evaluate (Section \ref{sec:subs}). Our more rigorous selection yielded a total sample of 305 RPE stars for dynamical analyses.

\section{Dynamical Properties} \label{subsec:dyn}

Estimated dynamical parameters for the VMP HK/HES (and RPE) sample have been obtained adopting the axisymmetric Galactic potential of \citet{mcmillan2017}. 
This model potential includes stellar thin and thick disks, gaseous disks, a flattened bulge, and spheroidal dark matter halo. In this model, the corresponding distance from the Sun to the Galactic center is $R_{\odot} = 8.2$ kpc \citep{Bland-Hawthorn, Gravity2019}.

\begin{figure}[ht!]
\centering
\includegraphics[height=19.8cm]{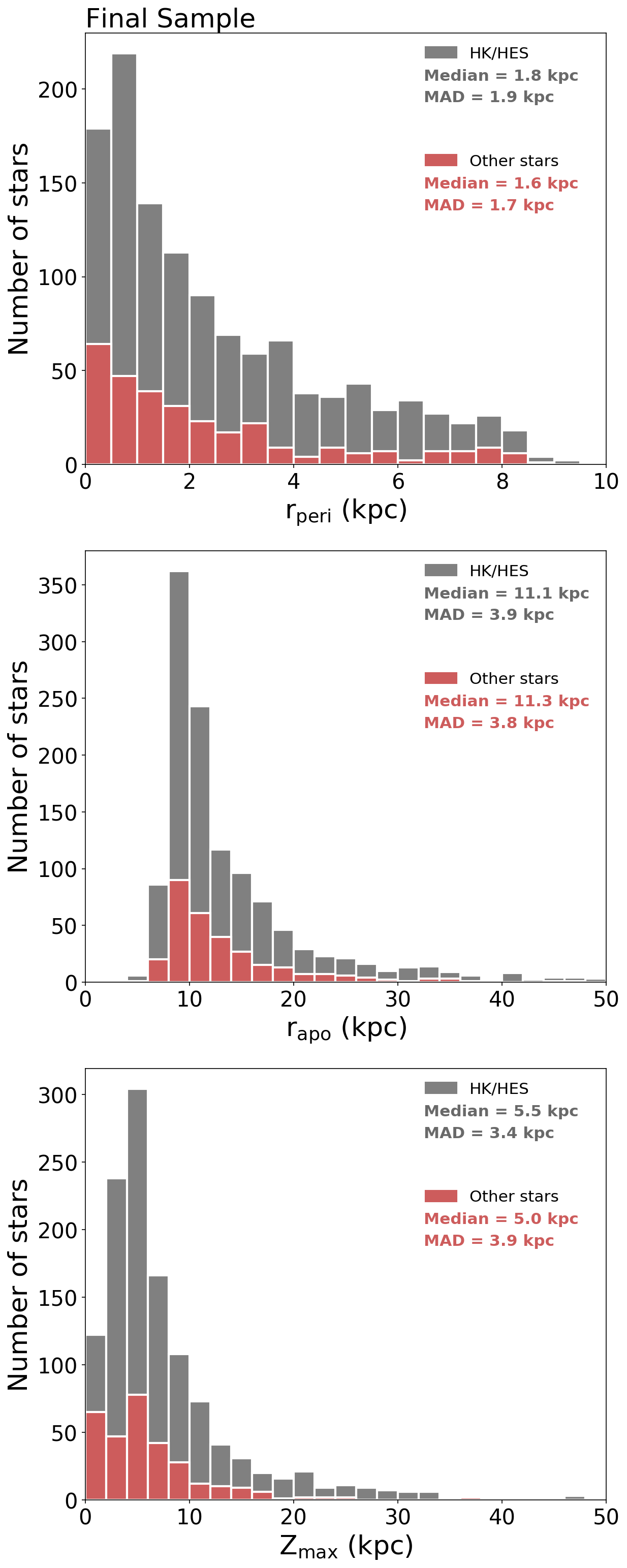}
\caption{ Histograms of the distributions of $r_{\rm peri}$ (top), $r_{\rm apo}$ (middle) and $Z_{\rm max}$ (bottom) for the Final Sample. Colors are the same as Figure \ref{fig:hists}. The $r_{\rm peri}$ is presented in 0.5 kpc bins, while the $r_{\rm apo}$ and $Z_{\rm max}$ are divided into 1.0 kpc bins. Medians and median absolute deviation (MAD) values have been listed following the color scheme of the histograms.}
\label{fig:hists_dyn}
\end{figure}

Stellar orbits have been integrated with the publicly available library \texttt{AGAMA} \citep{agama} to obtain estimates of the apocentric distance ($r_{\rm apo}$), pericentric distance ($r_{\rm peri}$), and eccentricity ($e = (r_{\rm apo} - r_{\rm peri}) / (r_{\rm apo} + r_{\rm peri})$) for each star (Figure \ref{fig:hists_dyn}). Energies ($E$) and actions ($J_R$, $J_{\phi}$, $J_z$; cylindrical coordinates) have also been computed with \texttt{AGAMA}, which implements the numerical method outlined by \citet{binney2012}. The actions\footnote{The complete formalism on actions can be found in Section 3.5 of \citet{BinneyBook}. For a practical interpretation of actions, see Figure 2 of \citet{trick2019}.} can be interpreted as follows. The radial action ($J_R \in  [0, \infty]$) is related to a star's orbital eccentricity, as it captures the extent of its radial excursion. The azimuthal action ($J_{\phi} \in [-\infty, \infty]$) represents the  stars' rotation around the Galactic center. Stars with  $J_{\phi} > 0$ are in prograde motion. The vertical action ($J_z \in [0, \infty]$) can be interpreted as the extent of the vertical excursion of a star's orbit relative to the Galactic plane. We have performed 1,000 Monte Carlo realizations for each star, taking into account uncertainties in the heliocentric distances, proper motions, and RVs, in order to assess their effect on the derived quantities.

The medians of the distributions of each dynamical quantity from the Monte Carlo realizations have been adopted as our nominal values. We have removed 14 stars presenting orbits that are formally unbound to the Galaxy ($E > 0$). Our Final Sample for the substructure search comprises 1,526 unique VMP stars with atmospheric parameters, six-dimensional phase-space vectors, energies, actions, and other dynamical quantities. The median metallicity of this final selected sample is [Fe/H] = $-2.3$, and its median absolute deviation (MAD) is 0.4 dex (Figure \ref{fig:hists}). 

Figure~\ref{fig:hists_dyn} shows histograms of the pericentric distances, $r_{\rm peri}$, apocentric distances, $r_{\rm apo}$, and the maximum distance from the Galactic plane achieved during the stars' orbits, $Z_{\rm max}$, for the stars in the Final Sample. From inspection of this figure, the distributions of these quantities for the HK/HES stars and the other stars (mostly calibration stars) are quite similar, justifying our choice to combine them for the clustering analysis.

\section{substructure search} \label{sec:subs}

We seek DTGs in the VMP HK/HES sample in the space defined by the energy and actions ($E$, $J_R$, $J_{\phi}$, $J_z$). These parameters (with slight variations) have been extensively used to search for the dynamical signatures of accreted material in the Galaxy  \citep{Helmi2017,roederer2018, myeongShards, koppelman2019,yuan2019,yuan2020b, yuan2020a, hansen2019,gudin2020a,naidu2020}.

\renewcommand{\arraystretch}{1.0}
\setlength{\tabcolsep}{0.50em}

\begin{table}[ht!]
\centering
\caption{DTGs in the VMP HK/HES Sample  
}
\label{tab:all_groups}
\begin{tabular}{cccl}
\hline
DTG & Members & Confidence & Comments  \\
\hline
\hline
1     & 9  & 84\%              & Sequoia  \\
2     & 8  & 87\%              & Polar, New$^\dag$           \\
3     & 18 & 88\%              & Helmi Stream  \\
4     & 8  & 66\%              & Polar, New  \\
5     & 10 & 66\%              & Sequoia      \\
6     & 6  & 80\%              & Rg5       \\
7     & 7  & 50\%              & Retrograde, New     \\
8     & 14 & 56\%              & ZY20:DTG-35    \\
9     & 7  & 58\%              & Polar, New  \\
10    & 5  & 94\%              & Prograde, New \\
11    & 13 & 67\%              & GSE \\
12    & 6  & 39\%              & Retrograde, New  \\
13    & 6  & 52\%              & ZY20:DTG-39 \\
14    & 11 & 66\%              & Retrograde, New            \\
15    & 6  & 80\%              & Prograde, New           \\
16    & 5  & 41\%              & ZY20:DTG-33                 \\
17    & 18 & 67\%              & Prograde, New \\ 
18    & 8  & 63\%              & ZY20:DTG-33           \\
19    & 6  & 51\%              & Retrograde, New  \\ 
20    & 6  & 57\%              & GSE  \\
21    & 8  & 62\%              & GSE           \\
22    & 7  & 62\%              & ZY20:DTG-33  \\
23    & 7  & 39\%              & GSE \\ 
24    & 12 & 62\%              & GSE  \\
25    & 10 & 41\%              & Thamnos  \\
26    & 8  & 59\%              & Thamnos          \\
27    & 13 & 88\%              & ZY20:DTG-19    \\
28    & 5  & 60\%              & GSE         \\
29    & 14 & 66\%              & GSE          \\
30    & 20 & 53\%              & GSE \\ 
31    & 12 & 47\%              & Thamnos          \\
32    & 13 & 70\%              & Thamnos \\ 
33    & 10 & 45\%              & Thamnos          \\
34    & 19 & 48\%              & GSE \\ 
35    & 8  & 57\%              & GSE          \\
36    & 21 & 64\%              & GSE          \\
37    & 30 & 61\%              & GSE \\ 
38    & 10 & 62\%              & GSE \\   \bottomrule \vspace{1px}
\end{tabular}
\begin{flushleft}
\ $^\dag$ Tentative association with other reported substructure (Section \ref{subsec:polar}). 
\end{flushleft}
\end{table}

\subsection{Clustering Method}

\label{subsec:clusters}

The identification of DTGs was carried out using the clustering algorithm Hierarchical
Density-Based Spatial Clustering of Applications with Noise  (\texttt{HDBSCAN}\footnote{\url{https://hdbscan.readthedocs.io}.}; \citealt{campello2015}), implemented in Python by \citet{hdbscan}.  

\texttt{HDBSCAN} has been developed to work with an initially unknown total number of groups, having variable shapes and density contrasts. Another important feature is that \texttt{HDBSCAN} is robust in the presence of noisy data; there is no dependence on the underlying assumption of smooth background models in energy-action space that past works relied upon to deal with this challenge. It is also convenient that the fundamental hyper-parameter that \texttt{HDBSCAN} requires to operate is the minimum number of elements to form a valid group (\texttt{min\_cluster\_size}), which is physically meaningful. 

\texttt{HDBSCAN} constructs  a hierarchical cluster tree based on the estimated local densities for each point in the multi-dimensional parameter space. Two points are considered connected if they form a dense region in this parameter space. The clusters are likely to be real if they are persistent for different density thresholds, from very low to very high. These can be interpreted as ``persistent clusters" (with a \texttt{min\_cluster\_size}) with less-likely members falling out of them as they move through the hierarchical tree. This exercise ensures that the resulting groups are very stable.

We have chosen \texttt{min\_cluster\_size} = 5 and \texttt{cluster\_selection\_method} = ``\texttt{leaf}" as input parameters in order to build the cluster hierarchy tree. This choice, in particular the ``\texttt{leaf}" mode, is optimized for the detection of fine-grained substructures in favor of larger ones. After testing with the algorithm, we have noticed that the overall behavior of the groups do not change significantly with \texttt{min\_cluster\_size}. Essentially, smaller clumps are erased when this parameter is increased, as expected. Therefore, we have avoided smaller values for the \texttt{min\_cluster\_size}, as that could lead to an unrealistic number of groups. We have also avoided values that are too large, as that would not be in keeping with our objective of finding small groups that could have originated from low-mass systems. The larger substructures in our data can still be mapped out following the execution of the procedure, assembled from the smaller, robust ones (Section \ref{sec:large}). This choice is also consistent with the work of both \citet{yuan2020a} and \citet{myeongStreamsAndClumps}, who accepted groups containing at least 4 and 5 members, respectively, making our study comparable to theirs. Again, we note that the clusters are built exclusively from the VMP sample. The RPE stars are mapped onto the groups only after the cluster assignment has already been completed.

Although we expect clusters identified by \texttt{HDBSCAN} to be quite stable, we still need to assess the statistical significance of the groups against variations in the dynamical properties of each member star due to their estimated uncertainties. We have sampled 1,000 sets of ($E$, $J_R$, $J_{\phi}$, $J_z$) from the 16th and 84th percentiles of each quantity for each star in a Monte Carlo framework. Then, we throw these ``perturbed" data sets back into the hierarchical tree, and re-evaluate their cluster assignments. An object is considered a valid member of a given group if it was assigned at least 200 times to the same cluster out of the 1,000 Monte Carlo realizations. We take this to indicate that the star presents at least a 20\% membership probability. We define the ``confidence" level of a given group (see Table \ref{tab:all_groups}) as the average membership probability of its member stars.

\begin{figure*}[ht!]
\centering
\includegraphics[width=18.0cm,height=17.0cm]{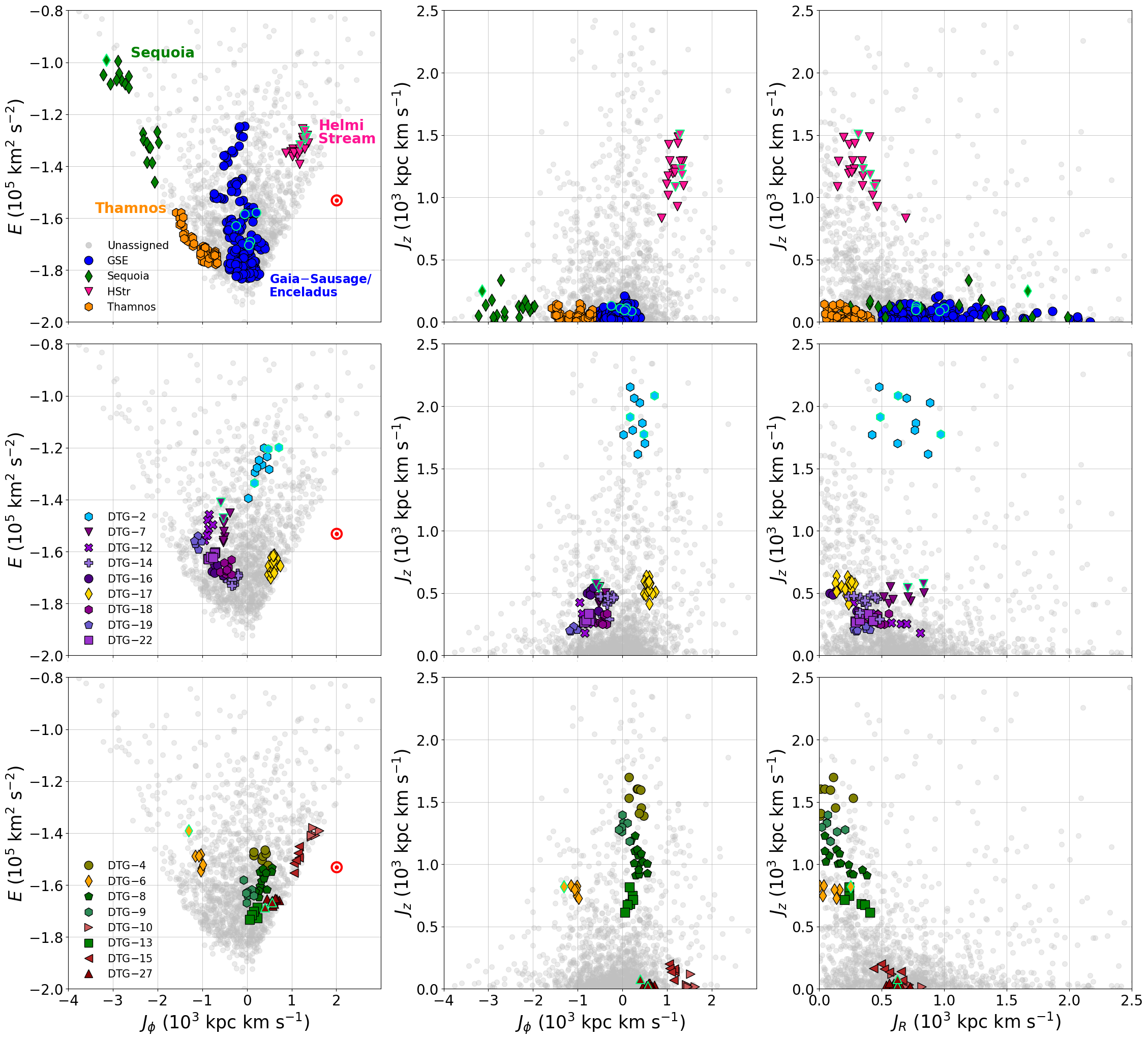}
\caption{Energy-action space plots of the VMP HK/HES sample. Left column: ($E$, $J_{\phi}$). The prograde, low-energy corner of the ($E$, $J_{\phi}$) diagrams are depopulated due to the $|V-V_{\rm{LSR}}|>210$\,km\,s$^{-1}$ criterion (Section \ref{subsec:hkhesVMPsample}). \textbf{The position of the Sun is marked with a red circle}. Middle column: ($J_z$, $J_{\phi}$). Right column: ($J_z$, $J_R$). The larger known substructures in the Galactic halo (Section \ref{sec:large}) that have been recognized in our sample are featured in the top row. Blue circles, dark green diamonds, pink triangles, and orange hexagons are stars associated to GSE, Sequoia, HStr, and Thamnos, respectively. The rest of the DTGs, including many new ones (see Table \ref{tab:new_groups}), are shown in the middle and bottom rows. DTGs with similar colors, but different symbols, have qualitatively similar dynamical properties (Section \ref{sec:new}).  In all of the plots, gray dots represent stars that were not found to be dynamically clustered.  The symbols with lime-colored edges are RPE stars associated with the different groups according to their colors and symbols (Section \ref{subsec:$r$-process}).}
\label{fig:ExJ}
\end{figure*}

Application of this method results in the identification of 38 significant DTGs\footnote{We follow the nomenclature proposed by \citet{yuan2020a}. Dynamical groups resulting from different analyses can be recognized by the initials of the first author's names, the year of publication, and the number of the DTG. Our DTG-1, for instance, would be referenced as GL20:DTG-1.}, comprising $\sim$400 stars (27\% of the Final Sample). The characteristics of each DTG are listed in Table \ref{tab:all_groups}, and are represented with different symbols in Figure \ref{fig:ExJ}. Those with qualitatively similar dynamical properties are shown with similar colors.  The number of stars in each dynamical group ranges from 5 to 30. The one with the highest confidence level is DTG-10 (94\%). The DTGs with the lowest confidence are DTG-12 and DTG-23, both at 39\%. Just over half (23/38; 60\%) are retrograde ($\langle J_{\phi} \rangle < 0$). We compare our results to those from the literature and further discuss the nature of our DTGs in Sections \ref{sec:large} and \ref{sec:new}.

\renewcommand{\arraystretch}{1.0}
\setlength{\tabcolsep}{0.35em}

\begin{table*}[ht!]
\centering
\caption{Larger Substructures in the VMP HK/HES Sample}
\label{tab:large_struc}
\begin{tabular}{@{}cccccccccc@{}}
\toprule
Substructure &
  Associated DTGs &
  Members &
  $\langle E \rangle$ &
  ($\langle J_R \rangle, \langle J_{\phi} \rangle, \langle J_z \rangle$) &
  $\langle e \rangle$ &
  $\langle i \rangle$ &
  ($\langle v_R \rangle$, $ \langle v_{\phi} \rangle$, $ \langle v_z \rangle$) &
  Median [Fe/H] \\
 &
   &
   &
   $\sigma_E$ &
  ($\sigma_{J_R}$, $\sigma_{J_{\phi}}$, $\sigma_{J_z}$) &
  $\sigma_e$ &
  $\sigma_i$ &
  ($\sigma_{v_R}$, $\sigma_{v_{\phi}}$, $\sigma_{v_z}$)
  & MAD$_{\rm [Fe/H]}$\\
      &                 &   & (km$^2$ s$^{-2}$)  & (kpc\,km\,s$^{-1}$)       &      & (deg) & (km\,s$^{-1}$) &       \\ \midrule \midrule
GSE   & 11,20,21,23,24,28,29,   & 173  & $-1.7 \times 10^5$ & ($826$,$-108$ , $57$)    & $0.90$ & $103$  & ($-7$, $-14$, $2$) & $-2.2$      \\
      & 30,34,35,36,37,38 &   & $1.4 \times 10^4$  & (318, 229, 39)   & 0.05 & 39  & (162, 28, 49) & $\; \; \; 0.3$  \\ \hline 
Sequoia  & 1,5               & 19    & $-1.2 \times 10^5$ & ($916$, $-2496$, $111$) & 0.54 & 163 & $0$, $-301$, $-39$) & $-2.2$  \\
      &                 &  & $1.5 \times 10^4$  & (576, 389, 73)    & 0.14 & 6   & (150, 44, 83) & $\; \; \; 0.4$ \\ \hline 
HStr & 3     & 18 & $-1.3 \times 10^5$ & ($321$, $1153$, $1210$) & 0.40 & 62 & ($-46$, $144$, $-198$) & $-2.3$   \\
      &                 &   & $3.3 \times 10^3$  & (132, 134, 183)     & 0.08 & 3  & (98, 16, 167) & $\; \; \; 0.5$ \\ \hline
Thamnos & 25,26,31,32,33     & 53 & $-1.7 \times 10^5$ & ($200$, $-1066$, $59$)   & 0.44 & 159 & ($-4$, $-136$, $6$) & $-2.2$     \\
      &                 &   & $5.7 \times 10^3$  & (99, 276, 41)      & 0.14 & 10  & (70, 36, 55) & $\; \; \; 0.3$ \\ \bottomrule \smallskip
\end{tabular}
\end{table*}

\section{Mapping Larger Substructures} \label{sec:large}
\subsection{GSE}
\label{subsubsec:gse}

The structure known as GSE has been suggested to be the remnant of the last large-scale merging event experienced by the Galaxy \citep{belokurov2018,Haywood2018, helmi2018}, containing the majority of the accreted stars in the nearby halo. Its members form a well-defined sequence in the color-magnitude diagram presented in \citet{gaiaHR} for stars with halo-like kinematics. They also exhibit typically low metallicities ([Fe/H] $\lesssim -0.7$; \citealt{DiMatteo2019}) and $\alpha$-element abundance ratios (\citealt{hayes2018, mack2019}; also noted earlier on by \citealt{nissen2010}). These stars are distinguishable in velocity space, as they form an extended distribution in $v_R$ (velocity towards the radial direction of the cylindrical coordinate system) around an azimuthal velocity ($v_{\phi}$) close to zero \citep{koppelman2018, feuillet2020}, which translates into highly eccentric orbits \citep{naidu2020}. Moreover, stars from the GSE progenitor are usually old ($\gtrsim$ 10 Gyr; \citealt{gallart2019, bonaca2020}), and its (proposed) globular clusters form a tight age-metallicity relation \citep{MyeongSausageClusters, myeongSequoia, massari2019, kraken}.

\begin{figure*}[ht!]
\centering
\includegraphics[scale=0.425]{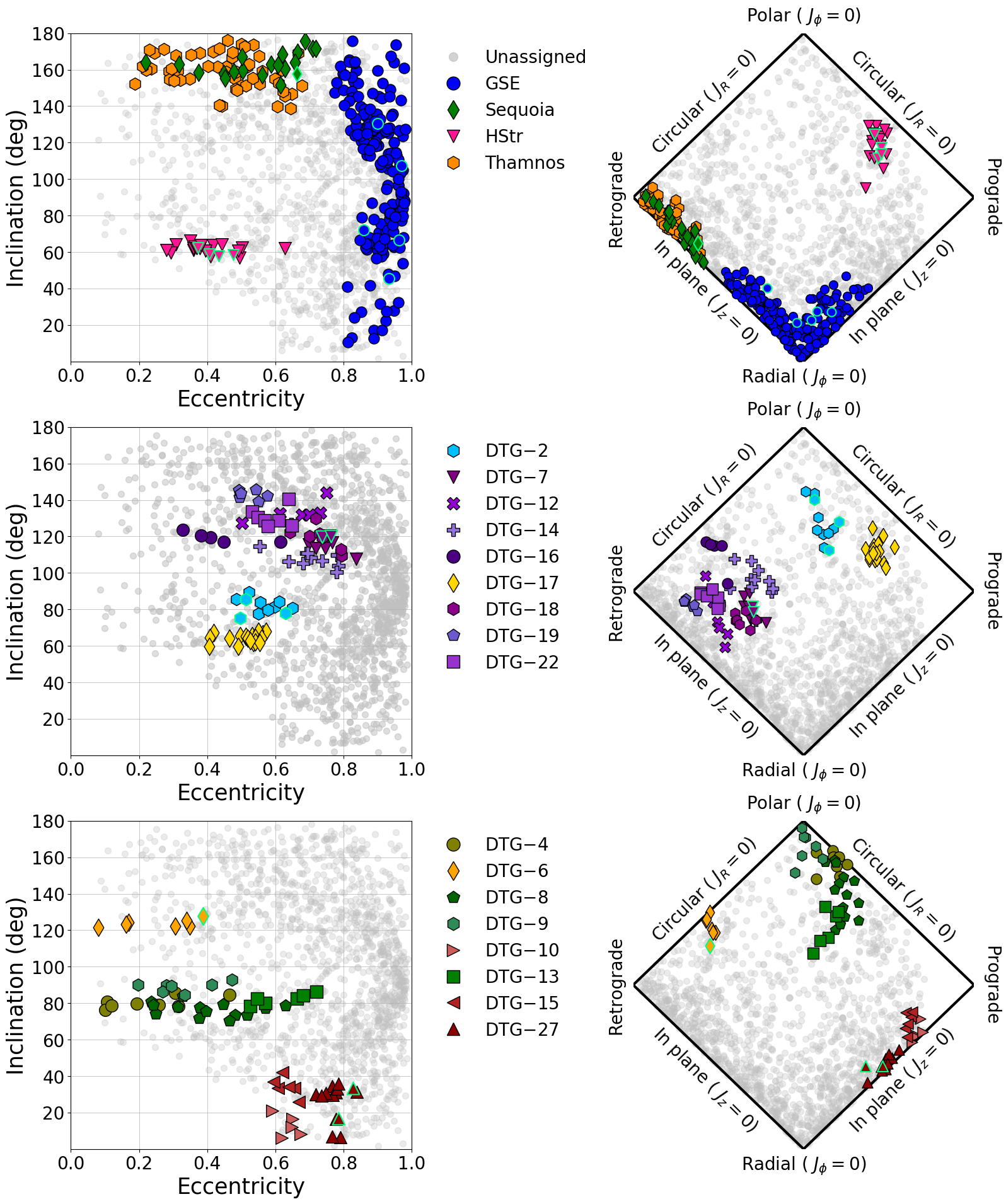}
\caption{Left column: Inclination versus eccentricity of the VMP HK/HES sample. Stars within $0^{\circ} < i < 90^{\circ}$ are in prograde ($J_{\phi} > 0$) motion, and those in the region of $90^{\circ} < i < 180^{\circ}$ are retrograde ($J_{\phi} < 0$). Right column: Projected action-space diagrams of the sample. The horizontal axis is $J_{\phi} / J_{\rm total}$, where $J_{\rm total} = J_R + |J_{\phi}| + J_z$. The vertical axis is $(J_z - J_R)/J_{\rm total}$. One can notice the absence of stars towards the prograde corner of the plots, due to the exclusion of objects from the disk system ($|V-V_{\rm{LSR}}|>210$\,km\,s$^{-1}$; Section \ref{subsec:hkhesVMPsample}). The DTGs found in the VMP sample are highlighted as the colored symbols (as in Figure \ref{fig:ExJ}). Symbols with lime-colored edges are RPE stars associated with the different groups according to their colors and symbols (Section \ref{subsec:$r$-process}). Gray dots represent unassigned stars.}
\label{fig:proj_act_1}
\end{figure*}

In order to identify potential members of GSE among our DTGs, we establish that these groups must display $\langle e \rangle \geq 0.8$ (see Figure \ref{fig:proj_act_1}). This criterion is similar to \citet{naidu2020} and \citealt{bonaca2020}. These authors have shown that stars in this range of eccentricity compose a well-behaved, strongly peaked metallicity distribution. The requirement that $e \geq 0.8$ had already been suggested by \citet{MyeongSausageClusters}, who considered the distribution of globular clusters and halo stars in action space, and \citet{mack2019}, who analyzed the $\alpha$-element abundance ratios of such high-eccentricity stars. In total, 13 of our DTGs (comprising 173 stars; Table \ref{tab:large_struc}) can be attributed to GSE (blue circles in the top row of Figure \ref{fig:ExJ}). Application of this selection yields a mean radial action $\langle J_R \rangle \gtrsim 450$\,kpc\,km\,s$^{-1}$ for each associated group. Additionally, our selected groups are contained within $-600 \lesssim \langle J_{\phi} \rangle$ (kpc\,km\,s$^{-1}$) $ < +500$ (similar to \citealt{feuillet2020}). The dynamical nature of GSE can be visualized in the projected action space diagram shown in Figure \ref{fig:proj_act_1} (top panel).

%We have included an extra requirement that a DTG must present $\langle J_z \rangle < 500$\,kpc\,km\,s$^{-1}$ (also adopted by \citealt{MyeongSausageClusters} and \citealt{yuan2020a}, making our work consistent with theirs) in order to be considered part of this larger substructure.} 

%Thirteen of our DTGs (comprising 173 stars; Table \ref{tab:large_struc}) can be attributed to GSE (blue circles in the top row of Figure \ref{fig:ExJ}). We consider those DTGs with $\langle J_z \rangle < 500$\,kpc\,km\,s$^{-1}$ and $\langle e \rangle \geq 0.7$ to be part of this larger substructure. Application of these selection criteria yields a mean radial action $\langle J_R \rangle \gtrsim 450$\,kpc\,km\,s$^{-1}$ for each associated group. The work of \citet{belokurov2018} also demonstrated that GSE can be characterized as having a mean azimuthal velocity ($v_{\phi}$) close to zero. In agreement with this, our selected groups are contained within $-600 \lesssim \langle J_{\phi} \rangle$ (kpc\,km\,s$^{-1}$) $ < +500$. The dynamical nature of GSE can be visualized in the projected action space diagram shown in Figure \ref{fig:proj_act_1} (top panel).

The GSE substructure, as we have defined it, presents stars with orbital inclinations (${i = \cos^{-1}{\left ( L_z/L \right )}}$, where $L_z$ is the vertical component of the total angular momentum, $L$) spanning all possible values. In the $i$ vs. $e$ space, it shows a ``boomerang-like" shape in the top panel of Figure \ref{fig:proj_act_1}, concentrated towards high values of eccentricity. This substructure presents characteristic $v_{\phi}$ and $v_R$ that overlap with those from the SD \citep{bonaca2017, DiMatteo2019, belokurov2020,an2020, amarante2020a,Amarante2020b}. However, the average metallicity of the SD is $\langle$[Fe/H]$\rangle$ $\approx -0.5$. Since our stars are all VMP, we expect minimal contamination from this source.

\begin{figure*}[ht!]
\centering
\includegraphics[width=18.0cm]{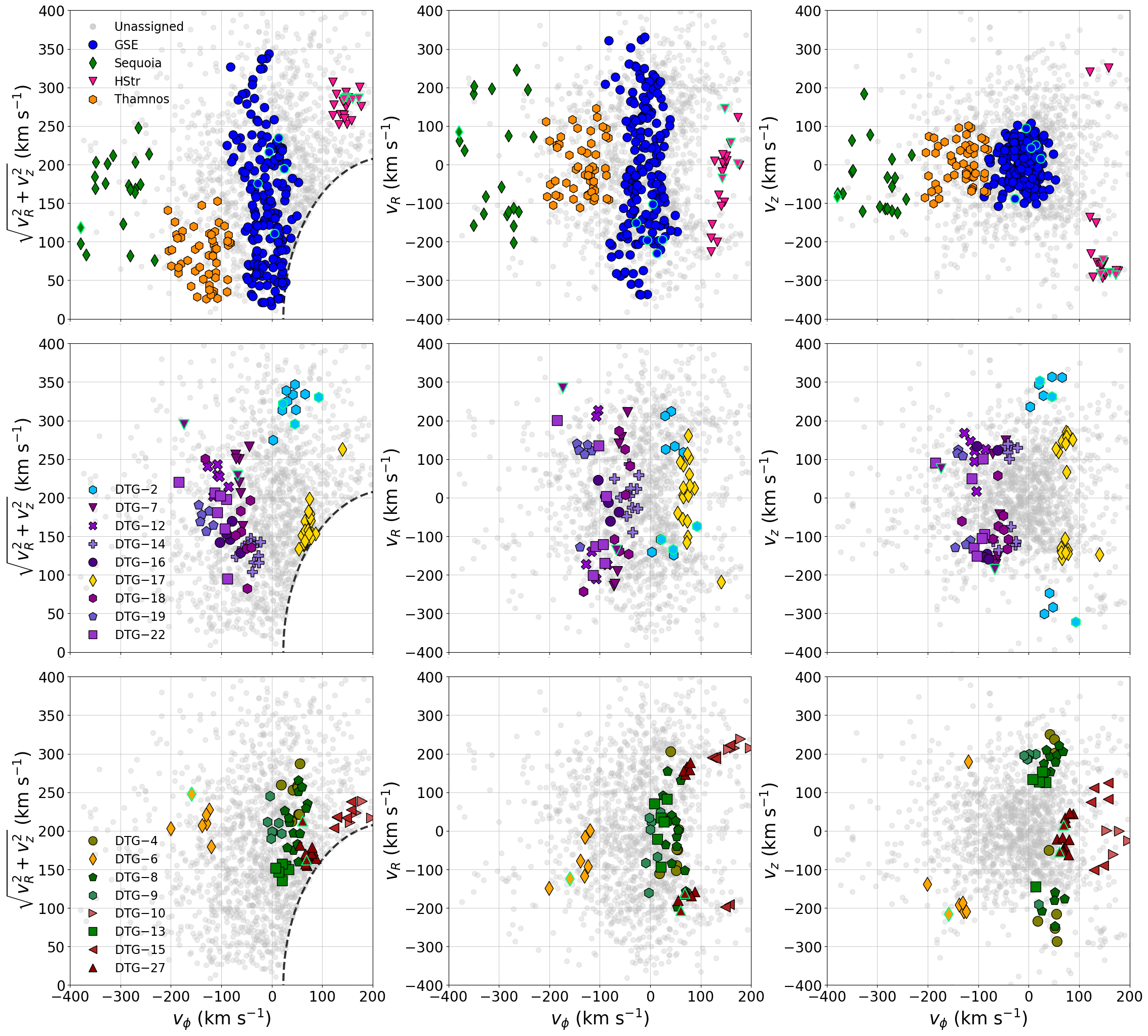}
\caption{Velocity-space diagrams of the VMP HK/HES sample. Left panels: ($\sqrt{v_R^2 + v_z^2}$, $v_{\phi}$). The dashed curve marks the selection boundary corresponding to the requirement that $|V-V_{\rm{LSR}}|>210$\,km\,s$^{-1}$ (Section \ref{subsec:hkhesVMPsample}). Middle panels: ($v_R$, $v_{\phi}$). Right panels: ($v_z$, $v_{\phi}$). Upper row: GSE (blue circles; Section \ref{subsubsec:gse}), Sequoia (dark green diamonds; Section \ref{subsubsec:seq}), the HStr (pink triangles; Section \ref{subsubsec:HStr}) and Thamnos (orange hexagons; Section \ref{subsubsec:tham}). Middle and bottom rows: The rest of the DTGs, including many new ones (Section \ref{sec:new}). The symbols with lime-colored edges are RPE stars associated with the different substructures according to their colors and symbols (Section \ref{subsec:$r$-process}). Gray dots represent unassigned stars.}
\label{fig:vels}
\end{figure*}

The velocity space of GSE is shown in Figure \ref{fig:vels}. As previously mentioned, this substructure presents an almost null net rotation with small dispersion ($\langle v_{\phi} \rangle \approx -14 $\,km\,s$^{-1}$; $\sigma_{v_{\phi}} = 28 $\,km\,s$^{-1}$). An interesting feature is that our velocity distribution in the radial direction is continuous, occupying $-300 < \langle v_R \rangle$ (km\,s$^{-1}$) $< +300$. This is very similar to the aforementioned characteristically huge spread in $v_R$ presented by \citet{belokurov2018} and others since (e.g., \citealt{koppelman2018} and \citealt{feuillet2020}). The importance of GSE to the halo is further examined in Section \ref{subsec:$r$-process}. 

\subsection{Sequoia}
\label{subsubsec:seq}

Speculation that a substantial, strongly retrograde merger event could have contributed stars to the Galactic halo began to appear many years ago \citep{norris1989, carollo2007, carollo2010, beers2012}. Further evidence came with the discovery of the massive globular cluster FSR 1758 \citep{cantat2018, barba2019}. \citet{myeongHalo} revealed an excess of stars with highly energetic, very retrograde orbits in the Galactic halo with metallicities of $-1.9 <$ [Fe/H] $< -1.3$. Building on that, \citet{myeongShards} discovered a profusion of small dynamical groups of stars in this region of the ($E$, $J_{\phi}$) space (and called them Rg1-4 and Rg6). 

Based on the above, \citet{myeongSequoia} argued that this population of stars originated from a merger event they referred to as Sequoia. These authors also claimed that a dynamically cohesive group of globular clusters (including FSR 1758) were associated to Rg1-4 and Rg6. So far, other independent analyses recognized additional Sequoia debris \citep{koppelman2019, massari2019, matsuno2019, dietz2020,yuan2020a,yuan2020b,monty2020,naidu2020}.

Among our dynamical groups, DTG-1 and DTG-5 (comprising a total of 19 stars, with a high confidence level, $\sim$75\%) can be readily identified as part of the Sequoia remnant. These DTGs have $\langle E \rangle = -1.2 \times 10^5$\,km$^2$\,s$^{-2}$ and $\langle J_{\phi} \rangle \approx -2500$\,kpc\,km\,s$^{-1}$, so they are well-within the interval established by \citet{myeongSequoia}. DTG-1 and DTG-5 also spread through many different values of $J_R$ (in line with \citealt{myeongShards}), and exhibit low average vertical action: $\langle J_{z} \rangle \approx 145$\,kpc\,km\,s$^{-1}$. We revisit the Sequoia event, and its possible connection to RPE stars, in Section \ref{subsec:$r$-process}.

\subsection{The Helmi Stream}
\label{subsubsec:HStr}

The Helmi Stream (HStr) was one of the first dynamical groups to be found. \citet{helmi1999} analyzed astrometric data of a small sample of metal-poor ([Fe/H] $<-1.6$) stars from the \textit{Hipparcos} Catalog \citep{perryman1997}, and found that some of these were significantly clumped in angular-momentum space. Further members were added by \citet{chiba2000} and other works since (e.g., \citealt{koppelman2018, myeongStreamsAndClumps}). One distinguishing aspect of this stream is that the majority of its members exhibit negative $v_z$, although some have positive $v_z$. This feature has been interpreted as the partially phase-mixed fragments of a (now) shredded satellite \citep{helmi2008}. The discovery of such a group, with apparently disconnected blobs in velocity space, is a demonstration of the power of searches in the integrals-of-motion space. More recent studies focused on the chemical aspects of this substructure \citep{roederer2010, aguado2020}, confirming its ancient nature from its low typical metallicity ($-3.0 \lesssim$ [Fe/H] $\lesssim -1.5$), its profile in [$\alpha$/Fe] (with a ``knee" at [Fe/H] $\approx -2.0$), and predominance of the $r$-process contribution to the enrichment of neutron-capture elements ([Sr/Ba] $\lesssim 0.0$) in the low-metallicity regime. However, the complicated story of the HStr and its progenitor is still under scrutiny (and debate; see, e.g., \citealt{koppelmanHelmi,naidu2020}), and its characterization is far from being complete.

Inspection of the first row of Figure \ref{fig:ExJ} allows for the immediate association of one of our larger (18 stars) and high-confidence (88\%) dynamical groups (DTG-3) with the HStr. This connection is in agreement with the selection criteria delineated by \citet{koppelmanHelmi}. Independent efforts converge on an azimuthal velocity of $\langle v_{\phi} \rangle \approx 150$\,km\,s$^{-1}$ for this stream \citep{beers2017,gaiaKinematics,koppelman2018,myeongStreamsAndClumps}, which is compatible with our findings ($\langle v_{\phi} \rangle = 144$\,km\,s$^{-1}$; $\sigma_{v_{\phi}} = 16 $\,km\,s$^{-1}$). Crucially, our DTG-3 has members that are detached in the vertical component of velocity (upper right panel of Figure \ref{fig:vels}), with the great majority presenting $v_z < 0$, which is expected for the canonical HStr. The potential enrichment of the HStr members in neutron-capture elements via the $r$-process is further explored in Section \ref{subsec:$r$-process}. \\

\subsection{Thamnos}
\label{subsubsec:tham}

\citet{koppelman2019} suggested that another substantial, very retrograde substructure exists in the Galactic halo, which they named Thamnos. It resides in the same corner region in the projected action-space map as Sequoia (Figure \ref{fig:proj_act_1}, top panel). However, there is a marked difference in orbital energy between Thamnos and Sequoia (upper left pane of Figure \ref{fig:ExJ}). The Thamnos event can be described, then, as a low-energy counterpart of Sequoia. These authors further suggested that this substructure could possibly be divided into two pieces, Thamnos 1 (Th. 1), highly retrograde and with lower metallicity, and Thamnos 2 (Th. 2), moderately retrograde and more metal rich.

From our analysis, a total of 5 dynamical groups could be associated with this substructure, 2 of them with Th. 1 (DTG-25 and DTG-26) and the other 3 with Th. 2 (DTG-31, DTG-32, and DTG-33). Since Th. 1 and Th. 2 have been proposed to share the same progenitor, we present them as a single cohesive substructure (orange hexagons; top row of Figure \ref{fig:ExJ}). Thamnos can be differentiated from GSE as having predominantly retrograde motion ($\langle i \rangle \approx 160^{\circ}$; $\langle J_{\phi} \rangle \approx -1000$\,kpc\,km\,s$^{-1}$) and lower eccentricity ($\langle e \rangle = 0.44$; $\langle J_R \rangle \approx 200$\,kpc\,km\,s$^{-1}$). These values are in agreement with the limits independently proposed by \citet{naidu2020}.

We note that the proliferation of small dynamical groups of stars in the low-energy, highly retrograde corner of the ($E$, $J_{\phi}$) space has also been pointed out by \citet{yuan2020a} in their own inspection of a VMP sample (their ZY20:DTG-21/24/29; see their Figure 6). Their DTGs are also comparable to ours in $J_R$ and $J_z$. It is possible that many (if not all) of the groups reported by both works are pieces of the same larger substructure. This hypothesis is attractive, since it has been shown that Thamnos has a metallicity distribution that preferentially occupies the VMP regime \citep{koppelman2019, helmi2020, naidu2020}. Thus, it would be natural to find it from examination of VMP stellar samples. 

Our findings corroborate the claims that Sequoia and Thamnos are well-separated in their binding energies, as can be seen from the upper left panel of Figure \ref{fig:ExJ}. If one were to assume that they originated from the same merger event, it would imply that the progenitor was at least as massive as that of GSE, in conflict with their lower metallicities. For the purpose of confirming the nature of these substructures, one requires not only reliable metallicities, but other chemical abundances, specially of the $\alpha$-elements, in order to explore their chemical-evolution and star-formation histories. It is clear from the ongoing discussions about Sequoia/Thamnos in the literature just how difficult it is to disentangle the formation history of the Galactic halo in terms of its clumps, streams, over-densities, and their respective progenitors.

\renewcommand{\arraystretch}{1.0}
\setlength{\tabcolsep}{0.50em}

\begin{table*}[htp!]
\centering
\caption{New DTGs in the VMP HK/HES Sample and their Likely Associations}
\label{tab:new_groups}
\begin{tabular}{ccccccccc}
\toprule
DTG &
  Association & Members &
  $\langle E \rangle$ &
  ($\langle J_R \rangle$,$ \langle J_{\phi} \rangle$,$ \langle J_z \rangle$) &
  $\langle e \rangle$ &
  $\langle i \rangle$ &
  ($\langle v_R \rangle$,$ \langle v_{\phi} \rangle$,$ \langle v_z \rangle$) &
  Median [Fe/H] \\
 & & & 
  $\sigma_E$ &
  ($\sigma_{J_R}$, $\sigma_{J_{\phi}}$, $\sigma_{J_z}$) &
  $\sigma_e$ &
  $\sigma_i$ &
  ($\sigma_{v_R}$, $\sigma_{v_{\phi}}$, $\sigma_{v_z}$) & 
  MAD$_{\rm [Fe/H]}$\\
   & & &  (km$^2$ s$^{-2}$)  & (kpc\,km\,s$^{-1}$)        &      & (deg) & (km\,s$^{-1}$) &        \\ \hline \hline
2 & New$^\dag$ & 8 & $-1.3 \times 10^5$ & (690, 293, 1878)  & 0.57 & 83 & (52, 35, 74) & $-2.4$ \\
 & & & $5.7 \times 10^3$ & (277, 189, 102) & 0.05 & 4  & (158, 19, 292) & $\; \; \; 0.4$    \\ \hline

4 & New & 8 & $-1.5 \times 10^5$ & (85, 330, 1535)  & 0.23 & 80 & ($-19$, 46, $-44$) & $-2.5$  \\
 & & & $2.8 \times 10^3$ & (89, 122, 110) & 0.13 & 3  & (104, 13, 237) & $\; \; \; 0.4$   \\

8 & ZY20:DTG-35 & 14  & $-1.6 \times 10^5$ & (173, 361, 1039)  & 0.40 & 77 & ($-11$, 48, $51$) & $-2.5$  \\
 & & & $3.7 \times 10^3$ & (110, 106, 99) & 0.12 & 4  & (102, 14, 192) & $\; \; \; 0.4$   \\

9 & New & 7  & $-1.6 \times 10^5$ & (93, 25, 1299)  & 0.32 & 89 & ($-28$, 3, $139$) & $-2.2$  \\
 & & & $2.7 \times 10^3$ & (63, 80, 66) & 0.09 & 3  & (77, 10, 145) & $\; \; \; 0.2$  \\

13 & ZY20:DTG-39 & 6  & $-1.7 \times 10^5$ & (299, 159, 709)  & 0.62 & 82 & ($16$, 20, $88$) & $-2.3$   \\
 & & & $1.7 \times 10^3$ & (82, 68, 70) & 0.08 & 3  & (65, 9, 114) & $\; \; \; 0.5$  \\ \hline 

6 & Rg5 & 6  & $-1.5 \times 10^5$ & (82, $-1046$, 787)  & 0.23 & 123 & ($-75$, $-140$, $-126$) & $-2.7$   \\
 & & & $2.7 \times 10^3$ & (67, 60, 41) & 0.11 & 2  & (58, 30, 152) & $\; \; \; 0.3$  \\ \hline

7 & New & 7  & $-1.5 \times 10^5$ & (644, $-500$, 471)  & 0.75 & 115 & ($-37$, $-62$, $-54$) & $-2.5$  \\
 & & & $4.2 \times 10^3$ & (118, 54, 45) & 0.04 & 4  & (200, 9, 134) & $\; \; \; 0.4$  \\

12 & New & 6  & $-1.5 \times 10^5$ & (576, $-868$, 285)  & 0.66 & 134 & ($-43$, $-108$, $112$)  & $-2.3$  \\
 & & & $3.7 \times 10^3$ & (191, 62, 84) & 0.09 & 6  & (204, 14, 53) & $\; \; \; 0.4$  \\

14 & New & 11  & $-1.7 \times 10^5$ & (370, $-326$, 431)  & 0.70 & 109 & ($6$, $-41$, $18$)  & $-2.4$  \\
 & & & $1.4 \times 10^3$ & (78, 101, 61) & 0.07 & 7  & (51, 14, 126) & $\; \; \; 0.4$  \\

16 & ZY20:DTG-33 & 5  & $-1.6 \times 10^5$ & (161, $-683$, 482)  & 0.44 & 120 & ($-12$, $-83$, $-40$)  & $-2.4$   \\
 & & & $1.8 \times 10^3$ & (84, 94, 73) & 0.11 & 3  & (41, 15, 154) & $\; \; \; 0.3$  \\

18 & ZY20:DTG-33 & 8  & $-1.7 \times 10^5$ & (473, $-450$, 278)  & 0.73 & 119 & ($-31$, $-64$, $-69$)  & $-2.2$  \\
 & & & $2.1 \times 10^3$ & (60, 82, 35) & 0.05 & 6 & (149, 30, 58)  & $\; \; \; 0.2$  \\

19  & New & 6 & $-1.6 \times 10^5$ & (339, $-1108$, 209)  & 0.53 & 143 & ($85$, $-133$, $-2$)  & $-2.4$  \\
 & & & $1.9 \times 10^3$ & (52, 58, 14) & 0.04 & 2  & (104, 11, 130)  & $\; \; \; 0.6$ \\

22 & ZY20:DTG-33 & 7  & $-1.7 \times 10^5$ & (353, $-780$, 297)  & 0.59 & 131 & ($-40$, $-111$, $-34$)  & $-2.2$   \\
 & & & $1.2 \times 10^3$ & (62, 65, 28) & 0.04 & 5 & (156, 34, 110) & $\; \; \; 0.4$  \\\hline

10  & New & 5  & $-1.4 \times 10^5$ & (701, 1496, 46)  & 0.64 & 13 & ($134$, 182, $-19$)  & $-2.1$  \\
 & & & $1.5 \times 10^3$ & (86, 80, 43) & 0.03 & 6  & (192, 20, 26) & $\; \; \; 0.2$  \\

15 & New & 6 & $-1.5 \times 10^5$ & (555, 1110, 146)  & 0.63 & 34 & ($68$, 140, $34$)  & $-2.2$  \\
 & & & $3.5 \times 10^3$ & (92, 55, 43) & 0.03 & 5 & (203, 15, 102) & $\; \; \; 0.2$  \\

27 & ZY20:DTG-19 & 13   & $-1.7 \times 10^5$ & (617, 590, 25)  & 0.77 & 26 & ($-16$, 73, $-17$) & $-2.2$  \\
 & & & $8.3 \times 10^2$ & (47, 62, 12) & 0.03 & 10  & (171, 8, 40) & $\; \; \; 0.2$  \\\hline 

17 & New & 18 & $-1.7 \times 10^5$ & (223, 577, 545)  & 0.51 & 64 & ($7$, 4, $0$) & $-2.3$  \\
 & & & $2.7 \times 10^3$ & (47, 63, 58) & 0.05 & 2 & (94, 18, 145) & $\; \; \; 0.4$  \\ \bottomrule \\

\end{tabular}
\vspace{-10px}
\begin{flushleft}
\ The DTGs are ordered by their numbers, but keeping them grouped according to their qualitatively similar dynamical properties, as in Section \ref{sec:new}. \\
\ $^\dag$ Tentative association with other reported substructure (Section \ref{subsec:polar}). 
\end{flushleft}
\end{table*}

\section{New Dynamical Groups}
\label{sec:new}

\subsection{New Polar Groups}
\label{subsec:polar}

Besides the HStr (DTG-3), other dynamical groups stand out from the rest as having predominantly polar orbits. One of these is DTG-2 (sky-blue hexagons; second row of Figure \ref{fig:ExJ}). This DTG has been segregated from the rest of the polar groups as it exhibits $\langle J_R \rangle > \langle J_{\phi} \rangle$; it has 8 members and a high confidence level (87\%). One of the distinctive features of DTG-2 can be appreciated from its distribution in velocity space (Figure \ref{fig:vels}). Much like the HStr, we notice that this group is split into two blobs of $v_z$, one positive and one negative. We note that this DTG exhibits the highest absolute value of vertical velocity ($\abs{v_z} \approx 300$\,km\,s$^{-1}$) out of our groups. This DTG might also be a fragmented piece of the low-mass stellar debris stream LMS-1 (Wukong), recently discovered by \citet{yuan2020b}. These authors also argued that this substructure would be reasonably metal-poor ([Fe/H] $\lesssim-1.5$), in keeping with our VMP selection. Further discussion on the nature of DTG-2 is presented in Section \ref{subsec:$r$-process}.

The remaining polar groups are DTG-4/8/9/13 (green symbols; bottom row of Figure \ref{fig:ExJ}). They are also mildly prograde ($v_{\phi} \approx +50$\,km\,s$^{-1}$; Figure \ref{fig:vels}), and present characteristically low radial actions ($\langle J_R \rangle \leq 300$\,kpc\,km\,s$^{-1}$; Figure \ref{fig:proj_act_1}). Out of these, DTG-8 and DTG-13 could be readily associated with ZY20:DTG-35 and ZY20:DTG-39 from \citet{yuan2020a}, respectively. These authors also identified many prograde polar groups. Future studies might reveal if this excess of polar VMP clumps could be the debris of a larger substructure or a superposition of individual small accreted systems. 

\subsection{New Prograde Groups}
\label{subsec:prograde}

Some of our DTGs have been classified as being predominantly prograde. Three of these are DTG-10/15/27 (red triangles; bottom row of Figure \ref{fig:ExJ}). The orbits of their member stars lie close to the Galactic plane ($\langle J_z \rangle \lesssim 150$\,kpc\,km\,s$^{-1}$; Figure \ref{fig:proj_act_1}), and exhibit moderate average eccentricities ($\langle e \rangle \geq 0.6$; Figure \ref{fig:proj_act_1}). All of these groups are also distinguishable from their kinematics; they present negative and positive blobs in $v_R$ (Figure \ref{fig:vels}). However, DTG-10 and DTG-15 are rotating much faster around the Galactic center ($\langle v_{\phi} \rangle \gtrsim +140$\,km\,s$^{-1}$; $\langle J_{\phi} \rangle \gtrsim +1100$\,kpc\,km\,s$^{-1}$) than DTG-27 ($\langle v_{\phi} \rangle \approx +70$\,km\,s$^{-1}$; $\langle J_{\phi} \rangle \approx +600$\,kpc\,km\,s$^{-1}$). The rotational motions of DTG-10/15 are compatible with the value suggested for the MWTD \citep{carollo2019,an2020}. Comparison with both \citet{carollo2014} and \citet{beers2014} also point to similarities in ($E$, $J_{\phi}$), but the Galactic gravitational-potential model used by these authors is different from ours, so this should be taken with caution. In addition, our kinematic cut (Section \ref{subsec:hkhesVMPsample}) should yield minimal contamination from stellar populations with disk-like orbits.

The more modest rotation of DTG-27 makes it overlap with both the SD and ZY20:DTG-19 \citep{yuan2020a} in $v_{\phi}$ and $v_R$. In Section \ref{subsubsec:gse}, we argued that stars from the SD should represent only a minor contamination in our VMP sample, since its metallicity is $\langle$[Fe/H]$\rangle \approx -0.5$. Indeed, these objects should not affect our definition of a large substructure such as GSE, but they could produce a small clump like DTG-27. The recent demonstration that a meaningful population of extremely and ultra metal-poor stars ([Fe/H] $\leq -3.0$ and $\leq -4.0$, respectively) permeates the Galactic thin and thick disks \citep{sestito2019, sestito2020, DiMatteo2020} underscores the possibility that these VMP stars could have acquired halo kinematics from dynamical heating mechanisms. The nature of DTG-27 is further discussed in Section \ref{subsec:$r$-process}.

The final prograde DTG out of our dynamical groups is DTG-17. Its stars are represented as yellow diamonds in the second row of Figure \ref{fig:ExJ}. This is a new group with 18 members (67\% confidence). Considering kinematics, this DTG is similar to the green polar groups (DTG-4/8/9/13; Section \ref{subsec:polar}). However, since its $\langle J_{\phi} \rangle > \langle J_z \rangle$, we present it as a predominantly prograde one. We note that \citet{yuan2020a} also found some DTGs in this region of the energy-action space (see their Figure 6). Detailed chemical abundances from future spectroscopic efforts might hint at the origin of this excess of prograde VMP dynamical groups.

\subsection{New Retrograde Groups}
\label{subsec:retrograde}

The majority of our new DTGs have retrograde orbits, many of which can be attributed to either Sequoia (Section \ref{subsubsec:seq}) or Thamnos (Section \ref{subsubsec:tham}). However, a total of 8 dynamical groups are apparently unrelated to any of these better known substructures. The first of these is DTG-6 (orange diamonds; bottom row of Figure \ref{fig:ExJ}). Its stars are very retrograde ($\langle J_{\phi} \rangle \approx -1000$\,kpc\,km\,s$^{-1}$), but exhibit moderate energy ($\langle E \rangle \approx -1.5 \times 10^5$\,km$^2$\,s$^{-2}$). This DTG occupies the same region as the Rg5 substructure proposed by \citet{myeongShards}, and is likely associated with it. A more 
in-depth examination of DTG-6/Rg5 is presented in Section \ref{subsec:$r$-process}.

The remaining retrograde DTGs are presented as purple symbols in the middle row of Figure \ref{fig:ExJ}. These are DTG-7/12/14/16/18/19/22. We note that this large set of DTGs occupies an intermediate region of the ($E$, $J_{\phi}$) space, between GSE and Thamnos. Therefore, it is difficult to disentangle their origins from dynamics alone. For instance, DTG-7 and DTG-12 have mean radial actions in the range of GSE ($\langle J_R \rangle \gtrsim 500$\,kpc\,km\,s$^{-1}$), but their large values of vertical and retrograde motions, respectively, makes them incompatible with this larger substructure. We revisit DTG-7 in Section \ref{subsec:$r$-process}. A similarly large set of VMP dynamical groups has already been reported by \citet{yuan2020a}. Among these, ZY20:DTG-33 strongly overlaps with three of our own DTGs (DTG-16, DTG-18, and DTG-22). Apparently, this region of the energy-action space is preferentially occupied by VMP stellar clumps; other searches, without [Fe/H] selection cuts, failed to recognize any cohesive substructures occupying it. Nevertheless, we are in urgent need for chemical-abundance information (e.g., $\alpha$-elements) for the members of these DTGs in order to test whether or not these might be the remnants of low-mass system(s) that merged into the Galaxy.

\section{Connections to RPE stars}
\label{subsec:$r$-process}

It has been argued, from a variety of standpoints, that DTGs of VMP stars are likely to be the debris of small systems shredded by the Galaxy in the past. Such low-mass (UFD and/or dSph) galaxies have also been suggested to be the probable environments in which $r$-process nucleosynthesis has yielded moderate and highly $r$-process enhancements \citep{ji2016, roed2016,roederer2018, hansen2017,gudin2020a}. Thus, it is useful to search for dynamical connections between RPE stars and the substructures identified in this work.

We have explored this possibility using the recent compilation of RPE stars from \citet{gudin2020a}, as described in Section \ref{subsec:rpeSample}. A final list of 305 stars with suitable dynamical parameters and [Eu/Fe] $> +0.3$ has been compiled, similar to the selection from \citet{roederer2018}, but expanded to include the $r$-I regime. We have performed dynamical calculations within the same scheme delineated in Section \ref{subsec:dyn}. Regarding the cluster assignments, we have carried out 1,000 Monte Carlo realizations of their ($E$, $J_R$, $J_{\phi}$, $J_z$), and fed these generated sets back into the cluster hierarchy tree (Section \ref{subsec:clusters}). This exercise is analogous to the one that has been applied to evaluate the statistical significance of our DTGs and estimate their confidence levels. Again, we have only considered stars with at least 20\% membership probability. We have retained stars that are both RPE and CEMP (CEMP-{\it{r}}), since the astrophysical site(s) for the production of this latter abundance pattern is still under investigation (see, e.g., \citealt{frebel2018}). 

A total of 18 RPE stars have been associated with our DTGs (Table \ref{tab:$r$-process}); some are members of previously known groups and others of our newly discovered ones. The locations of these stars in comparison to their respective DTGs can be appreciated in Figures \ref{fig:ExJ}, \ref{fig:proj_act_1}, and \ref{fig:vels}. They are represented with the same colors and symbols of their associated groups, but with lime-colored edges.

Out of these RPE stars, 5 of them have been associated to dynamical groups that belong to GSE. Two of these RPE stars have also been clustered together by \citet{gudin2020a} from their dynamics.
All of these stars are contained within the ranges $+0.30 <$ [Eu/Fe] $\lesssim +0.70$ and $-2.5 <$ [Fe/H] $< -2.0$. These [Eu/Fe] ratios are comparable to RPE stars from the Ursa Minor dSph galaxy \citep{ sadakane2004, cohen2010} for the same metallicity interval. However, stars from Ursa Minor present systematically lower abundance values of [Ba/Eu]. \citet{yuan2020a} had already hinted at a connection between RPE stars and GSE from their dynamics. The accumulated evidence points to GSE (and/or its progenitor systems) as an important source of such stars.

\renewcommand{\arraystretch}{1.0}
\setlength{\tabcolsep}{0.60em}

\begin{table*}[ht!]
\centering
\caption{Associations of the VMP HK/HES Dynamical Groups with Recognized RPE Stars}
\begin{tabular}{cccccccccc}
\toprule
DTG & Associations     & Star           & Confidence & {[}Fe/H{]} & {[}Eu/Fe{]} & {[}Ba/Eu{]} & {[}C/Fe{]$_c$} & Class \\ \midrule \midrule

1 & Sequoia & 2MASS J11444086$-$0409511 & 20\%       & $-2.52$    & $+0.58$  &  $-0.84$  &   $+0.33$ & $r$-I  \\ \hline

2 & New$^\dag$ & 2MASS J00453930$-$7457294 & 21\%       & $-2.00$    & $+0.55$  &  $-0.18$ & $+0.98$ & $r$-I / CEMP-r \\
  &            & 2MASS J00413026$-$4058547 & 46\%       & $-2.58$    & $+0.38$  &  $-0.65$  &   $+0.20$ & $r$-I \\
  &            & 2MASS J02274104$-$0519230 & 78\%       & $-2.38$    & $+0.42$  &  $-0.60$  &   $+0.08$ & $r$-I \\ \hline

3 & HStr    & HD 175305 & 100\%       & $-1.50$    & $+0.44$  &  $-0.32$  &   & $r$-I  \\
  &         & HD 119516 & 100\%       & $-2.26$    & $+0.34$  &  $-0.36$  &   $+0.28$ & $r$-I     \\
  &         & BD+30:2611 & 100\%       & $-1.40$    & $+0.45$  &  $-0.37$  &  & $r$-I    \\
  &      & 2MASS J03270229+0132322 & 66\% & $-2.39$ & $+1.07$  &  $-0.57$  &   $+0.36$ & $r$-II  \\ \hline

6 & Rg5    & SDSS J235718.91$-$005247.8 & 22\%      & $-3.36$    & $+1.92$  &  $-0.80$  &  $+0.43$ & $r$-II   \\ \hline

7 & New    & BPS BS 16089$-$0013      & 39\%       & $-2.70$    & $+0.46$  &  $-0.59$  &  $+0.66$ & $r$-I      \\
  &        & 2MASS J17060555+0412354 & 20\%       & $-2.71$    & $+0.50$  &  $-0.45$  &  $+0.60$ & $r$-I   \\ \hline

24 & GSE    & BPS CS 22968$-$0026        & 91\%       & $-2.57$    &        $+0.58$  &  $-1.09$  &  $-0.09$ & $r$-I   \\
   &        & 2MASS J00482431$-$1041309 & 21\%       & $-2.50$    &        $+0.45$  &  $-0.36$  &  $+0.47$ & $r$-I  \\
28 &        & 2MASS J08393460$-$2122069 & 55\%       & $-1.94$    &        $+0.42$  &  $-0.29$  &  $+0.14$ & $r$-I \\
30 &        & 2MASS J18562774$-$7251331 & 28\%       & $-2.26$    &        $+0.32$  &  $-0.33$  &  $+0.24$ & $r$-I \\
   &        & 2MASS J00073817$-$0345509 & 22\%       & $-2.09$    &        $+0.73$  &  $-0.62$  &  $+0.17$ & $r$-II \\ \hline

27 & ZY20:DTG-19 & HD 115444                 & 100\%       & $-2.99$    &  $+0.85$  &  $-0.67$  &  $+0.32$ & $r$-II \\
   &             & 2MASS J20005766$-$2541488 & 20\%       & $-2.05$    &   $+0.40$  &  $-0.05$  &  $+0.27$ & $r$-I \\ 

\bottomrule
\end{tabular}
\begin{flushleft}
\ The DTGs are ordered by their numbers, but keeping them grouped according to their dynamical properties (Section \ref{sec:new}) and including the larger structures (Section \ref{sec:large}). \\

 \ $^\dag$ Tentative association with other reported substructure (Section \ref{subsec:polar}).\\
 \ [C/Fe]$_c$ values have been corrected for their evolutionary status \citep{placcoCaronCorr2014}.
\end{flushleft}
\label{tab:$r$-process}
\end{table*}

From examination of retrograde groups in our VMP sample, a $r$-I star is possibly connected to the Sequoia remnant (DTG-1). Additionally, our DTG-6 (compatible with Rg5; \citealt{myeongShards}) has been shown to be potentially associated to one of the most extremely RPE stars ($r$-II; [Eu/Fe] $= +1.92$) known to date. This is in line with \citet{yuan2020a}, who had already assigned the same object to the Rg5 debris. The level of $r$-process enrichment of this star is comparable to the RPE UFD galaxy Reticulum II \citep{ji2016}. Likewise, DTG-7 appears associated with 2 RPE (both $r$-I) stars. The similarities in [Fe/H], [Eu/Fe], and [Ba/Eu] for these objects can be seen in Table \ref{tab:$r$-process}. Interestingly, these abundance ratios are also comparable to those of RPE stars associated with the GSE substructure.

Only one of our predominantly prograde groups, DTG-27, has been associated with 2 RPE stars, one $r$-I and the other $r$-II. This finding is in agreement with \citet{gudin2020a}, who independently attributed both of these stars to the same dynamical group. In Section \ref{subsec:prograde}, we pointed out that a group of VMP stars with very similar orbital properties to DTG-27 had already been found (ZY20:DTG-19; \citealt{yuan2020a}). Member stars of this DTG should be primary targets for future studies, as they could provide clues on the early nucleosynthesis processes that operated in this component of the Galactic halo system.

Among our polar groups, 3 RPE stars, all $r$-I (one being of the CEMP-{\it{r}} sub-class) have been attributed to DTG-2. This is even more interesting, considering the tentative connection between this dynamical group and the \citet{yuan2020b} low-mass stellar debris stream LMS-1 (see also \citealt{naidu2020}; Section \ref{subsec:polar}). Future chemical-abundance analysis would be useful to confirm this linkage. 

One of the most intriguing of our findings is that 4 RPE stars have been associated with the HStr (DTG-3). Three of these stars have very similar [Eu/Fe] ratios ([Eu/Fe] $\approx +0.40$; $r$-I), and the other is more enriched in $r$-process elements ([Eu/Fe] $=+1.07$; $r$-II). The excess of RPE stars in this substructure is also consistent with recent spectroscopic efforts \citep{roederer2010, aguado2020}. In both works, the authors argued that the enrichment in neutron-capture elements of the member stars of the HStr was dominated by the $r$-process, based on their [Sr/Ba] patterns at the VMP end. Clearly, more detailed elemental-abundance studies of stars from the HStr would be useful to better constrain the chemical-evolution history of its progenitor. \\

\section{Conclusions}
\label{sec:conclusion}

In this work, we have considered the dynamical properties of the very metal-poor (VMP; [Fe/H] $\lesssim -2.0$) stars primarily selected from the HK/HES surveys. We have employed a sample of 1,526 VMP stars from the Galactic halo with suitably accurate dynamical parameters to perform a substructure search in the energy-action space. This metallicity cut allows us to find groups of stars that have conceivably been born in ultra-faint (UFD) and dwarf spheroidal (dSph) galaxies that merged with the Milky Way in the past.

Our analysis has been carried out with the algorithm \texttt{HDBSCAN}, where the clustering has been performed in the parameter space of ($E$, $J_R$, $J_{\phi}$, $J_z$). We have identified 38 significant Dynamically Tagged Groups (DTGs), comprising $\sim$400 stars. We have been able to recover larger, previously known substructures such as \textit{Gaia}-Sausage/Enceladus (GSE; \citealt{belokurov2018, helmi2018}), Sequoia \citep{myeongSequoia}, the Helmi Stream (HStr; \citealt{helmi1999}), and Thamnos \citep{koppelman2019}, as well as smaller groups from \citet{myeongShards} and \citet{yuan2020a}, and a number of newly identified ones. We have also investigated possible connections between our DTGs and $r$-process-enhanced (RPE) stars compiled by \citet{gudin2020a}. In total, 18 such stars have been associated to our groups, including several of the new ones. The main results from our analysis are summarized below.\\

$\bullet$ All of the aforementioned larger structures present meaningful numbers of VMP stars (at least $\sim$20 each) within the HK/HES sample. Future high-precision abundance analyses of these objects might allow us to constrain the conditions of the star forming environments and chemical evolution of their progenitors.

$\bullet$ The GSE substructure is associated with 173 of our stars in 13 DTGs. We have also provided evidence that 5 $r$-I/$r$-II stars could be associated with it. This is in agreement with \citet{gudin2020a}; these authors also dynamically clustered together 2 of these stars, in an independent analysis. These results indicate that the GSE progenitor(s) might have been an important source of RPE stars to the Galactic halo.

$\bullet$ The distribution of our DTGs in the energy-action space favors the hypothesis that Sequoia and Thamnos are indeed distinct entities, being widely separated in energy, in line with other recent results \citep{koppelman2019, naidu2020,monty2020}. Also, DTG-1 (Sequoia) has been associated with one $r$-I star.

$\bullet$ The HStr has been recovered, in agreement with the literature \citep{koppelman2018, koppelmanHelmi, myeongStreamsAndClumps}. It has also been associated with 4 RPE stars. This could be another strong indication that, in the very low-metallicity regime, the progenitor of the HStr experienced enrichment in neutron-capture elements predominantly via the $r$-process, in accord with other recent results from chemical-abundance analyses \citep{roederer2010, aguado2020}.

$\bullet$ Some of our DTGs have highly polar orbits. One of these (DTG-2) has been tentatively associated with the low-mass stellar stream LMS-1 (or Wukong; see \citealt{naidu2020}) recently discovered by \citet{yuan2020b}. Surprisingly, 3 RPE stars have been attributed to this DTG. The rest of our polar DTGs are mildly prograde, and some are compatible with dynamical groups from \citet{yuan2020a}.

$\bullet$ Four of our DTGs have been classified as predominantly prograde. One of them, DTG-27, is associated with 2 RPE stars, in agreement with \citet{gudin2020a}. \citet{yuan2020a} identified a dynamical group with similar dynamical properties (YZ20:DTG-19). Its location in the energy-action and velocity spaces suggest that this DTG could be related to the dynamically heated disk of the Galaxy.

$\bullet$ Many of our smaller groups are strongly retrograde and have moderate energies. One of them (DTG-6) is comparable to Rg5 \citep{myeongShards} in both the energy-action and velocity spaces. One of the most extremely RPE stars known has been associated to this DTG; \citet{yuan2020a} also argued that this $r$-II star is potentially associated with Rg5. Seven other DTGs present predominantly retrograde orbits. Among these, DTG-7 has been associated with 2 $r$-I stars with very similar [Eu/Fe] and [Ba/Eu] ratios; its member stars are clearly compelling targets for future studies. \\

The complex formation history of the Galactic stellar halo has been a long-standing mystery. However, large astrometric, photometric, and spectroscopic surveys have now provided the tools for Galactic Archaeologists to start solving this puzzle. Ancient very metal-poor stars play a crucial role in reconstructing the formation history of the Milky Way's halo, and provide clues on the environments in which the early nucleosynthesis of heavy elements took place. The VMP field stars in the HK/HES (and other surveys) have the enormous advantage that they are much closer (and hence significantly brighter) than any surviving satellite galaxy. Thus, the methodology described in this work can be applied to future (and much larger) stellar samples with more complete chemical-abundance information, helping to unveil and refine our understanding of the assembly history of the Milky Way. 

\acknowledgments

We thank the anonymous referee for her/his useful comments and suggestions that helped improve the manuscript. GL acknowledges CAPES (PROEX) for the funding of his Ph.D (Proc. 88887.481172/2020-00). SR would like to acknowledge partial financial support from FAPESP (Proc. 2015/50374-0 and 2014/18100-4), CAPES, and CNPq. GL, SR, TCB, HDP, APV, RMS, YA, VMP, and AF acknowledge partial support from grant PHY 14-30152, Physics Frontier Center/JINA Center for the Evolution of the Elements (JINA-CEE), awarded by the US National Science Foundation. HDP thanks FAPESP Proc. 2018/21250-9. APV acknowledges FAPESP for the postdoctoral fellowship No. 2017/15893-1 and the DGAPA-PAPIIT grant IG100319. RMS acknowledges CNPq (Project 436696/2018-5). YSL acknowledges support from the National Research Foundation (NRF) of Korea grant funded by the Ministry of Science and ICT (No. 2017R1A5A1070354 and NRF-2018R1A2B6003961). NC acknowledges funding by the Deutsche Forschungsgemeinschaft (DFG, German Research Foundation) -- Project-ID 138713538 -- SFB 881 (``The Milky Way System'', subproject A04). JR acknowledges support from the AAS Small Research Grant, awarded by the American Astronomical Society (AAS) in 2001 and 2002. AF acknowledges partial support from NSF grant AST-1716251. This work has made use of data from the European Space Agency (ESA) mission {\it Gaia} (\url{https://www.cosmos.esa.int/gaia}), processed by the {\it Gaia} Data Processing and Analysis Consortium (DPAC,
\url{https://www.cosmos.esa.int/web/gaia/dpac/consortium}). Funding for the DPAC has been provided by national institutions, in particular the institutions participating in the {\it Gaia} Multilateral Agreement. This research has made use of the SIMBAD database, operated at CDS, Strasbourg, France. George Preston and Stephen Shectman are recognized for the original conception of an objective-prism search for very metal-poor halo stars at a time when their existence in any great numbers was still very much in question. Dieter Reimers and Lutz Wisotzki are recognized for their initiation of the Hamburg/ESO survey and development of the automated scanning techniques for the HES objective-prism plates. TCB and NC extend heartfelt thanks to all of the observers, telescope facilities, support staff, and time-allocation committees who enabled the hundreds of nights of photometric and spectroscopic follow-up of candidate metal-poor stars from the HK survey and the HES over the past three decades.

\newpage

\facilities{AAT 3.9m, CTIO 1.5m, CTIO 4m, ESO Danish 1.5m, ESO 1.5m, ESO 3.6m, ESO NTT, GEM-N 8.1m, GEM-S 8.1m, INT 2.5m, KPNO 0.9m, 2.1m, KPNO 4m, LCO 2.5m, OHP 1.5m, PAL 5m, SOAR 4.1m, SAAO 0.7m, 1.7m, SSO 2.3m, UKS 1.2m.}


\software{
          {\tt matplotlib} \citep{matplotlib},
          {\tt Numpy} \citep{numpy},
          {\tt R-project} \citep{Rproject},
          {\tt scipy} \citep{scipy},
           {\tt pandas} \citep{pandas}.
          }

\clearpage

\bibliographystyle{aasjournal}

\bibliography{bibliography.bib}

\appendix

\section{Initial and Final Samples of VMP Stars from the HK/HES Surveys}

\subsection{Initial Sample}
\label{sec:append_Init}
Table \ref{tab:initial_sample} provides the relevant information for the Initial Sample (Section \ref{sec:data}), as well as their derived atmospheric parameters, including metallicity (in the form of [Fe/H]) and carbonicity ([C/Fe]). We note that a small fraction ($\sim$3\%) of these stars are quite cool and carbon enhanced, with $T_{\rm eff} \leq 4500$\,K and [C/Fe] (and/or [C/Fe]$_c$) $\geq +0.7$. Caution is urged when considering the listed values of [Fe/H], [C/Fe], and [C/Fe]$_c$ for these stars, due to the presence of ``carbon-veiling" in cooler CEMP stars, which depresses the continuum in the region of the Ca~II K line that has a dominant influence on the metallicity estimation (see discussion in \citealt{yoon2020}). These effects can be mitigated by application of procedures similar to those described in Yoon et al., which are currently being refined, and will be employed for the present data in the near future.  

The first column of the table lists the  2MASS \citep{2MASS} names, when available.  The second and third columns list the HK and HES names. Other names for stars that do not appear in 2MASS or the HK and/or HES surveys are provided in the fourth column. The $V$ magnitude is listed as well, based on information reported in the literature, supplemented, where needed, with transformed values from \textit{Gaia} DR2 photometry \citep{GaiaDR2Photometry}, or estimated from the HES prism plates.

%%%%%%%%%%%%%%%%%%%%%%%%%%%%%%%%%%%%%%%%%%%%%%%%%%%%%

\renewcommand{\arraystretch}{1.0}
\setlength{\tabcolsep}{0.35em}

%\begin{longrotatetable}
\begin{longtable}{>{\small}c >{\small}c >{\small}c >{\small}c >{\small}c >{\small}r >{\small}c >{\small}c >{\small}c >{\small}l >{\small}r >{\small}r}
\caption{Initial Sample of the VMP HK/HES Stars} \\
\toprule
Name & Name & Name  & Name & RA & \multicolumn{1}{>{\small}c}{DEC}& $V$ mag& $T_{\rm eff}$ & $\log g$  & \multicolumn{1}{>{\small}c}{[Fe/H]} & \multicolumn{1}{>{\small}c}{[C/Fe]} & \multicolumn{1}{>{\small}c}{[C/Fe]$_c$} \\
(2MASS) & (HK) & (HES) & (other) & (deg) & \multicolumn{1}{>{\small}c}{(deg)} &  & (K)  & (cgs) & \multicolumn{1}{>{\small}c}{} & \multicolumn{1}{>{\small}c}{} \\ \midrule \midrule
\endfirsthead
\multicolumn{12}{c}%
{{\bfseries Table \thetable\ }\textit{(continued)}} \\
\toprule
Name & Name & Name  & Name & RA & \multicolumn{1}{>{\small}c}{DEC}& $V$& $T_{\rm eff}$ & $\log g$  & \multicolumn{1}{>{\small}c}{[Fe/H]} & \multicolumn{1}{>{\small}c}{[C/Fe]} & \multicolumn{1}{>{\small}c}{[C/Fe]$_c$} \\
(2MASS) & (HK) & (HES) & (other) & (deg) & \multicolumn{1}{>{\small}c}{(deg)} &  & (K)  & (cgs) & \multicolumn{1}{>{\small}c}{} & \multicolumn{1}{c}{} \\ \midrule \midrule
\endhead
\hline
\multicolumn{12}{c}{\textit{(continued)}}\\
\endfoot
\endlastfoot
00000093$-$3037362 & $\dots$ & HE 2357$-$3054 & $\dots$  & 0.0039  & $-$30.6267 & 16.3 & 4358 & 0.75 & $-$3.50$^\dagger$ & +3.38$^\dagger$ & +3.42$^\dagger$ \\
00001345$-$2705322 & $\dots$ & HE 2357$-$2722 & $\dots$  & 0.0560  & $-$27.0923 & 16.8 & 6128 & 3.30 & $-$1.90   & $-$0.08 & $-$0.08 \\
00003365$-$4158196 & $\dots$ & HE 2357$-$4215 & $\dots$  & 0.1402  & $-$41.9721 & 16.6 & 6630 & 3.73 & $-$2.48   & +1.11 & +1.11 \\
00003955$-$1622127 & CS 29517$-$0037 & $\dots$ & $\dots$  & 0.1649  & $-$16.3702 & 14.7 & 6168 & 3.49 & $-$2.21   & +1.16 & +1.16 \\
00004942$-$2914458 & CS 22961$-$0019 & $\dots$ & $\dots$  & 0.2060  & $-$29.2461 & 14.1 & 5752 & 4.18 & $-$1.88   & +0.89 & +0.89 \\
00024896$-$1834447 & CS 30304$-$0033 & $\dots$ & $\dots$  & 0.2800  & $-$18.5790 & 13.6 & 6544 & 3.67 & $-$2.44   & $<$+1.11 & $<$+1.11 \\
00011421$-$2230539 & $\dots$ & HE 2358$-$2247 & $\dots$  & 0.3092  & $-$22.5150 & 16.1 & 6251 & 3.65 & $-$2.07   & +1.06 & +1.06 \\
00012661$-$0036118 & $\dots$ & HE 2358$-$0052 & $\dots$  & 0.3609  & $-$0.6033  & 16.4 & 6330 & 3.43 & $-$1.81   & +1.04 & +1.04 \\
00014549$-$0549465 & CS 22957$-$0022 & $\dots$ & $\dots$  & 0.4395  & $-$5.8296  & 13.3 & 5467 & 3.02 & $-$2.85   & +0.79 & +0.79 \\
00014710$-$1347251 & CS 31060$-$0062 & $\dots$ & $\dots$  & 0.4463  & $-$13.7903 & 14.0 & 6257 & 3.36 & $-$1.84   & $<$$-$0.06 & $<$$-$0.06 \\
\bottomrule
\multicolumn{12}{l}{ [C/Fe]$_c$ values have been corrected for their evolutionary status \citep{placcoCaronCorr2014}.} \\
\multicolumn{12}{l}{ $^\dagger$ CEMP stars with $T_{\rm eff} \leq 4500$\,K. Their [Fe/H], [C/Fe], and [C/Fe]$_c$ ratios should be taken with caution (Section \ref{sec:append_Init}).} \\
\multicolumn{12}{l}{ This table is available in its entirety in machine-readable form.}
\label{tab:initial_sample}
\end{longtable}

\vfill

\subsection{Final Sample}
\label{sec:append_Final}
Table \ref{tab:final_sample} provides the stars contained in the Final Sample (Section \ref{sec:data}). The measured RVs (corrected to the heliocentric frame) from the medium-resolution spectroscopy are listed, along with the \textit{Gaia} DR2 values, where available. The \texttt{StarHorse} distance estimates and their relative errors are from \citet{anders2019}, as described in Section \ref{subsec:hkhesVMPsample}. Proper motions are taken from \citet{gaiadr2}. The primary derived dynamical properties used in our analysis (Section \ref{subsec:dyn}) are also listed.

%%%%%%%%%%%%%%%%%%%%%%%%%%%%%%%%%%%%%%%%%%%%%%%%%%%%%%%%%%%%%%%%%%

%\newpage\phantom{blabla}

\renewcommand{\arraystretch}{1.0}
\setlength{\tabcolsep}{0.30em}

%\begin{longrotatetable}

\begin{longtable}{>{\scriptsize}c >{\scriptsize}r >{\scriptsize}c >{\scriptsize}c >{\scriptsize}r >{\scriptsize}r >{\scriptsize}c >{\scriptsize}c >{\scriptsize}c >{\scriptsize}c >{\scriptsize}c}
\caption{Final Sample of the VMP HK/HES Stars} \\
\toprule
Name &
  \multicolumn{1}{>{\scriptsize}c}{RV} &
  \multicolumn{1}{>{\scriptsize}c}{RV$_{Gaia}$} &
  D (err) &
  \multicolumn{1}{>{\scriptsize}c}{PM$_{\rm RA}$} &
  \multicolumn{1}{>{\scriptsize}c}{PM$_{\rm DEC}$} &
  $E$ &
  ($J_R$, $J_{\phi}$, $J_z$) &
  Ecc &
  i &
  ($v_R$, $v_{\phi}$, $v_z$) \\
& & & & & & \multicolumn{1}{>{\scriptsize}c}{$\times 10^5$} &  \multicolumn{1}{>{\scriptsize}c}{$\times 10^3$} & & & \\ 
(2MASS) &
  \multicolumn{1}{>{\scriptsize}c}{(km\,s$^{-1}$)} &
  \multicolumn{1}{>{\scriptsize}c}{(km\,s$^{-1}$)} &
  (kpc) &
  \multicolumn{1}{>{\scriptsize}c}{(mas yr$^{-1}$)} &
  \multicolumn{1}{>{\scriptsize}c}{(mas yr$^{-1}$)} &
  (km$^2$ s$^{-2}$) &
  (kpc\,km\,s$^{-1}$) &
  & (deg) &
  (km\,s$^{-1}$) \\ \midrule \midrule
\endfirsthead
\multicolumn{11}{c}%
{{\bfseries Table \thetable\ } \textit{(continued)}} \\
\toprule
Name &
  \multicolumn{1}{>{\scriptsize}c}{RV} &
  \multicolumn{1}{>{\scriptsize}c}{RV$_{Gaia}$} &
  D (err) &
  \multicolumn{1}{>{\scriptsize}c}{PM$_{\rm RA}$} &
  \multicolumn{1}{>{\scriptsize}c}{PM$_{\rm DEC}$} &
  $E$ &
  ($J_R$, $J_{\phi}$, $J_z$) &
  Ecc &
  i &
  ($v_R$, $v_{\phi}$, $v_z$) \\
& & & & & & \multicolumn{1}{>{\scriptsize}c}{$\times 10^5$} & \multicolumn{1}{>{\scriptsize}c}{$\times 10^3$} & & \\ 
(2MASS) &
  \multicolumn{1}{>{\scriptsize}c}{(km\,s$^{-1}$)} &
  \multicolumn{1}{>{\scriptsize}c}{(km\,s$^{-1}$)} &
  (kpc) &
  \multicolumn{1}{>{\scriptsize}c}{(mas yr$^{-1}$)} &
  \multicolumn{1}{>{\scriptsize}c}{(mas yr$^{-1}$)} &
  (km$^2$ s$^{-2}$) &
  (kpc\,km\,s$^{-1}$) &
    & (deg) &
  (km\,s$^{-1}$) \\ \midrule \midrule
\endhead
\hline
        \multicolumn{11}{c}{\textit{(continued)}} \\
        \endfoot
        \endlastfoot
00003955$-$1622127 & $-$126     & $\dots$        & 1.87 (0.27)  & 20.15     & $-$28.54     & $-$1.74 & (0.24, $-$0.85, 0.09)  & 0.53  & 154.9       & (29.14, $-$105.03, 38.38)     \\
00014549$-$0549465 & $-$32      & $\dots$        & 3.33 (0.55)  & 8.20      & $-$26.45     & $-$1.40 & (0.13, $-$1.38, 0.87)  & 0.28  & 127.2       & ($-$139.43, $-$165.20, $-$153.78) \\
00020802$-$0249121 & $-$135     & $-$122.6   & 0.83 (0.07)  & $-$37.55    & $-$91.35     & $-$1.25 & (2.07, $-$0.17, 0.02)  & 0.97  & 123.5       & ($-$336.41, $-$21.02, 2.83)     \\
00023699$-$6545212 & 160      & $\dots$        & 6.30 (1.03)  & $-$2.90     & $-$6.21      & $-$1.53 & (0.55, $-$0.37, 0.66)  & 0.72  & 110.1       & ($-$210.65, $-$58.73, 23.58)    \\
00025685$-$4101045 & 123      & $\dots$        & 1.46 (0.15)  & 0.77      & $-$44.29     & $-$1.66 & (0.59, $-$0.47, 0.11)  & 0.79  & 130.4       & ($-$168.76, $-$60.25, $-$38.09)   \\
00030319$-$4020529 & 124      & $\dots$        & 0.78 (0.04)  & $-$34.77    & $-$96.56     & $-$1.35 & (1.63, $-$0.31, 0.03)  & 0.94  & 136.6       & ($-$310.42, $-$38.31, $-$7.06)    \\
00031324$-$0249074 & $-$10      & $\dots$        & 1.27 (0.17)  & 27.98     & $-$44.47     & $-$1.70 & (0.24, $-$0.49, 0.48)  & 0.55  & 113.1       & (1.81, $-$59.24, $-$134.86)     \\
00033134$-$0444224 & $-$100     & $\dots$        & 1.20 (0.15)  & 38.64     & $-$22.27     & $-$1.73 & (0.74, $-$0.15, 0.03)  & 0.94  & 131.4       & (123.78, $-$17.59, 3.10)      \\
\end{longtable}
%\end{longrotatetable}

%%%%%%%%%%%%%%%%%%%%%%%%%%%%%%%%%%%%%%%%%%%%%%%%%%%%%%%%%%%%%%%%%
\newpage\phantom{blabla}

\renewcommand{\arraystretch}{1.0}
\setlength{\tabcolsep}{0.3em}

%\begin{longrotatetable}
\begin{longtable}{>{\footnotesize}c >{\footnotesize}c >{\footnotesize}c >{\footnotesize}c >{\footnotesize}c >{\footnotesize}r >{\footnotesize}c >{\footnotesize}c >{\footnotesize}c >{\footnotesize}l >{\footnotesize}r >{\footnotesize}r}
\caption{Initial Sample of the VMP HK/HES Stars} \\
\toprule
Name & Name & Name  & Name & RA & \multicolumn{1}{>{\footnotesize}c}{DEC}& $V$ mag& $T_{\rm eff}$ & $\log g$  & \multicolumn{1}{>{\footnotesize}c}{[Fe/H]} & \multicolumn{1}{>{\footnotesize}c}{[C/Fe]} & \multicolumn{1}{>{\footnotesize}c}{[C/Fe]$_c$} \\
(2MASS) & (HK) & (HES) & (other) & (deg) & \multicolumn{1}{>{\footnotesize}c}{(deg)} &  & (K)  & (cgs) & \multicolumn{1}{>{\footnotesize}c}{} & \multicolumn{1}{>{\footnotesize}c}{} \\ \midrule \midrule
\endfirsthead
\multicolumn{12}{c}%
{{\bfseries Table \thetable\ }\textit{(continued)}} \\
\toprule
Name & Name & Name  & Name & RA & \multicolumn{1}{>{\footnotesize}c}{DEC}& $V$ mag& $T_{\rm eff}$ & $\log g$  & \multicolumn{1}{>{\footnotesize}c}{[Fe/H]} & \multicolumn{1}{>{\footnotesize}c}{[C/Fe]} & \multicolumn{1}{>{\footnotesize}c}{[C/Fe]$_c$} \\
(2MASS) & (HK) & (HES) & (other) & (deg) & \multicolumn{1}{>{\footnotesize}c}{(deg)} &  & (K)  & (cgs) & \multicolumn{1}{>{\footnotesize}c}{} & \multicolumn{1}{c}{} \\ \midrule \midrule
\endhead
\hline
\multicolumn{12}{c}{\textit{(continued)}}\\
\endfoot
\endlastfoot
00000093$-$3037362 & $\dots$ & HE 2357$-$3054 & $\dots$  & 0.0039  & $-$30.6267 & 16.3 & 4358 & 0.75 & $-$3.50$^\dagger$ & +3.38$^\dagger$ & +3.42$^\dagger$ \\
00001345$-$2705322 & $\dots$ & HE 2357$-$2722 & $\dots$  & 0.0560  & $-$27.0923 & 16.8 & 6128 & 3.30 & $-$1.90   & $-$0.08 & $-$0.08 \\
00003365$-$4158196 & $\dots$ & HE 2357$-$4215 & $\dots$  & 0.1402  & $-$41.9721 & 16.6 & 6630 & 3.73 & $-$2.48   & +1.11 & +1.11 \\
00003955$-$1622127 & CS 29517$-$0037 & $\dots$ & $\dots$  & 0.1649  & $-$16.3702 & 14.7 & 6168 & 3.49 & $-$2.21   & +1.16 & +1.16 \\
00004942$-$2914458 & CS 22961$-$0019 & $\dots$ & $\dots$  & 0.2060  & $-$29.2461 & 14.1 & 5752 & 4.18 & $-$1.88   & +0.89 & +0.89 \\
00024896$-$1834447 & CS 30304$-$0033 & $\dots$ & $\dots$  & 0.2800  & $-$18.5790 & 13.6 & 6544 & 3.67 & $-$2.44   & $<$+1.11 & $<$+1.11 \\
00011421$-$2230539 & $\dots$ & HE 2358$-$2247 & $\dots$  & 0.3092  & $-$22.5150 & 16.1 & 6251 & 3.65 & $-$2.07   & +1.06 & +1.06 \\
00012661$-$0036118 & $\dots$ & HE 2358$-$0052 & $\dots$  & 0.3609  & $-$0.6033  & 16.4 & 6330 & 3.43 & $-$1.81   & +1.04 & +1.04 \\
\end{longtable}

%%%%%%%%%%%%%%%%%%%%%%%%%%%%%%%%%%%%%%%%%%%%%%%%%%%%%%%%%%%%%%%%%%

\newpage\phantom{blabla}

\renewcommand{\arraystretch}{1.0}
\setlength{\tabcolsep}{0.30em}

%\begin{longrotatetable}

\begin{longtable}{>{\scriptsize}c >{\scriptsize}r >{\scriptsize}c >{\scriptsize}c >{\scriptsize}r >{\scriptsize}r >{\scriptsize}c >{\scriptsize}c >{\scriptsize}c >{\scriptsize}c >{\scriptsize}c}
%\begin{deluxetabletable*}{crrcrrccccc}
\caption{Final Sample of the VMP HK/HES Stars} \\
\toprule
Name &
  \multicolumn{1}{>{\scriptsize}c}{RV} &
  \multicolumn{1}{>{\scriptsize}c}{RV$_{Gaia}$} &
  D (err) &
  \multicolumn{1}{>{\scriptsize}c}{PM$_{\rm RA}$} &
  \multicolumn{1}{>{\scriptsize}c}{PM$_{\rm DEC}$} &
  $E$ &
  ($J_R$, $J_{\phi}$, $J_z$) &
  Ecc &
  i &
  ($v_R$, $v_{\phi}$, $v_z$) \\
& & & & & & \multicolumn{1}{>{\scriptsize}c}{$\times 10^5$} &  \multicolumn{1}{>{\scriptsize}c}{$\times 10^3$} & & & \\ 
(2MASS) &
  \multicolumn{1}{>{\scriptsize}c}{(km\,s$^{-1}$)} &
  \multicolumn{1}{>{\scriptsize}c}{(km\,s$^{-1}$)} &
  (kpc) &
  \multicolumn{1}{>{\scriptsize}c}{(mas yr$^{-1}$)} &
  \multicolumn{1}{>{\scriptsize}c}{(mas yr$^{-1}$)} &
  (km$^2$ s$^{-2}$) &
  (kpc\,km\,s$^{-1}$) &
  & (deg) &
  (km\,s$^{-1}$) \\ \midrule \midrule
\endfirsthead
\multicolumn{11}{c}%
{{\bfseries Table \thetable\ } \textit{(continued)}} \\
\toprule
Name &
  \multicolumn{1}{>{\scriptsize}c}{RV} &
  \multicolumn{1}{>{\scriptsize}c}{RV$_{Gaia}$} &
  D (err) &
  \multicolumn{1}{>{\scriptsize}c}{PM$_{\rm RA}$} &
  \multicolumn{1}{>{\scriptsize}c}{PM$_{\rm DEC}$} &
  $E$ &
  ($J_R$, $J_{\phi}$, $J_z$) &
  Ecc &
  i &
  ($v_R$, $v_{\phi}$, $v_z$) \\
& & & & & & \multicolumn{1}{>{\scriptsize}c}{$\times 10^5$} & \multicolumn{1}{>{\scriptsize}c}{$\times 10^3$} & & \\ 
(2MASS) &
  \multicolumn{1}{>{\scriptsize}c}{(km\,s$^{-1}$)} &
  \multicolumn{1}{>{\scriptsize}c}{(km\,s$^{-1}$)} &
  (kpc) &
  \multicolumn{1}{>{\scriptsize}c}{(mas yr$^{-1}$)} &
  \multicolumn{1}{>{\scriptsize}c}{(mas yr$^{-1}$)} &
  (km$^2$ s$^{-2}$) &
  (kpc\,km\,s$^{-1}$) &
    & (deg) &
  (km\,s$^{-1}$) \\ \midrule \midrule
\endhead
\hline
        \multicolumn{11}{c}{\textit{(continued)}} \\
        \endfoot
        \endlastfoot
00003955$-$1622127 & $-$126     & $\dots$        & 1.87 (0.27)  & 20.15     & $-$28.54     & $-$1.74 & (0.24, $-$0.85, 0.09)  & 0.53  & 154.9       & (29.14, $-$105.03, 38.38)     \\
00014549$-$0549465 & $-$32      & $\dots$        & 3.33 (0.55)  & 8.20      & $-$26.45     & $-$1.40 & (0.13, $-$1.38, 0.87)  & 0.28  & 127.2       & ($-$139.43, $-$165.20, $-$153.78) \\
00020802$-$0249121 & $-$135     & $-$122.6   & 0.83 (0.07)  & $-$37.55    & $-$91.35     & $-$1.25 & (2.07, $-$0.17, 0.02)  & 0.97  & 123.5       & ($-$336.41, $-$21.02, 2.83)     \\
00023699$-$6545212 & 160      & $\dots$        & 6.30 (1.03)  & $-$2.90     & $-$6.21      & $-$1.53 & (0.55, $-$0.37, 0.66)  & 0.72  & 110.1       & ($-$210.65, $-$58.73, 23.58)    \\
00025685$-$4101045 & 123      & $\dots$        & 1.46 (0.15)  & 0.77      & $-$44.29     & $-$1.66 & (0.59, $-$0.47, 0.11)  & 0.79  & 130.4       & ($-$168.76, $-$60.25, $-$38.09)   \\
00030319$-$4020529 & 124      & $\dots$        & 0.78 (0.04)  & $-$34.77    & $-$96.56     & $-$1.35 & (1.63, $-$0.31, 0.03)  & 0.94  & 136.6       & ($-$310.42, $-$38.31, $-$7.06)    \\
00031324$-$0249074 & $-$10      & $\dots$        & 1.27 (0.17)  & 27.98     & $-$44.47     & $-$1.70 & (0.24, $-$0.49, 0.48)  & 0.55  & 113.1       & (1.81, $-$59.24, $-$134.86)     \\
00033134$-$0444224 & $-$100     & $\dots$        & 1.20 (0.15)  & 38.64     & $-$22.27     & $-$1.73 & (0.74, $-$0.15, 0.03)  & 0.94  & 131.4       & (123.78, $-$17.59, 3.10)      \\
\bottomrule
\end{longtable}

\end{document}